\documentclass[aps,prd,preprint,nofootinbib]{revtex4-1}
\usepackage{amsfonts}
\usepackage{mathrsfs}
\usepackage{graphicx}
\usepackage{amsmath}
\usepackage{amssymb}
\usepackage{subcaption}
\usepackage{epsfig}
\usepackage{graphicx}
\usepackage{slashed}
\usepackage{color}
\usepackage{multirow}
\usepackage{ulem}
\usepackage{adjustbox}
\usepackage{stmaryrd}
\newcommand{\bqa}{\begin{eqnarray}}
	\newcommand{\eqa}{\end{eqnarray}}
\newcommand{\beq}{\begin{equation}}
	\newcommand{\eeq}{\end{equation}}
\def\non{\nonumber}

\def\B{{\cal B}}

\allowdisplaybreaks[1]

\graphicspath{{fig/}{dia/}} \DeclareGraphicsExtensions{.eps}
\usepackage[figuresleft]{rotating}

\begin{document}
	
	\title{Study of singly heavy baryon lifetimes}
	\author {Hai-Yang Cheng$^1$\footnote{phcheng@phys.sinica.edu.tw} and  Chia-Wei Liu$^2$\footnote{ron1a2l3d4@gmail.com}}
	\affiliation{	
		$^1$Institute of Physics, Academia Sinica,	Taipei, Taiwan 115, Republic of China\\
		$^2$School of Fundamental Physics and Mathematical Sciences, Hangzhou Institute for Advanced Study, UCAS, Hangzhou 310024, China
	}
	
	\date{\today}

\begin{abstract}
We study the inclusive decay widths of singly heavy baryons with the improved bag model in which the unwanted center-of-mass motion is removed.
Additional insight is gained by comparing the charmed and bottom baryons. 
We discuss the running of the baryon matrix elements and compare the results with the non-relativistic quark model~(NRQM). 
While the calculated  two-quark operator elements are compatible with the literature,  those of the four-quark ones deviate largely.
In particular,
the heavy quark limit holds reasonably well in the bag model for four-quark operator matrix elements but is badly broken in the NRQM.
We predict $1-\tau(\Omega_b)/ \tau(\Lambda_b^0) = (8.34\pm2.22)\%$ in accordance with the current experimental value of $(11.5^{+12.2}_{-11.6})\%$ 
and compatible
with $(13.2\pm 4.7)\%$ obtained in the NRQM. 
We find an excellent agreement between theory and experiment for the lifetimes of bottom baryons.
We confirm that
 $\Omega_c^0$ could live longer than $\Lambda_c^+$ 
after the dimension-7 four-quark 
operators are taken into account.
We recommend to measure
some semileptonic inclusive branching fractions in the forthcoming experiments to discern different approaches.  For example, we obtain ${\cal BF} (\Xi_c^+ \to X e^+ \nu_e) =
(8.57\pm 0.49)\% 
$ and 
${\cal BF} (\Omega_c^0 \to X e^+ \nu_e) =
(1.88\pm 1.69)\% 
$ in sharp contrast to 
${\cal BF} (\Xi_c^+ \to X e^+ \nu_e) =
(12.74^{+2.54}_{-2.45})\% 
$ and 
${\cal BF} (\Omega_c^0 \to X e^+ \nu_e) =
(7.59^{+2.49}_{-2.24})\% 
$ found in the NRQM. 
\end{abstract}

	\maketitle
	
	\section{Introduction}

The study of lifetimes for ground-state singly charmed and bottom baryons, both experimentally and theoretically, are of great interest. On the experimental side, the current world averages of bottom baryon lifetimes given by (in units of ps) \cite{PDG2022}
\begin{eqnarray} \label{eq:exptlifetime}
&& \tau(\Lambda^0_b)= 1.471\pm 0.009, \qquad
\tau(\Xi^0_b)= 1.480\pm0.030,
  \nonumber \\
&& \tau(\Xi^-_b)= 1.572\pm0.040, \qquad
\tau(\Omega^-_b)= 1.64^{+0.18}_{-0.17}
\end{eqnarray}
indicate the lifetime hierarchy 
\begin{equation}
\tau(\Xi_b^-)>\tau(\Xi_b^0)\simeq\tau(\Lambda_b^0),
\end{equation}
where the uncertainty of the $\Omega_b^-$ lifetime is too large to draw a conclusion yet.

On the contrary, the measured lifetimes of $\Omega_c^0$ and $\Xi_c^0$ have been drastically changed in 2018 and 2019, respectively, especially for the former (see Table \ref{tab:expt_lifetimes}). According to the 2004 version of the Particle Data Group (PDG), the lifetime pattern of charmed baryons was given by \cite{PDG2004}
\begin{equation} \label{eq:Bclifetimehierarchy}
\tau(\Xi_c^+)>\tau(\Lambda_c^+)>\tau(\Xi_c^0)>\tau(\Omega_c^0),
\end{equation}
  where $\Omega_c^0$ is  shortest-lived in the charmed baryon system.  
This lifetime hierarchy  remained stable from 2004 till 2018 \cite{PDG2018}. However, the situation was dramatically changed in 2018 when LHCb reported a new measurement of the charmed baryon $\Omega_c^0$ lifetime using semileptonic $b$-hadron decays \cite{LHCb:tauOmegac}. The new $\Omega_c^0$ lifetime $\tau(\Omega_c^0)=(268\pm24\pm10\pm2)\,{\rm fs}$ obtained by LHCb is nearly four times larger than the 2018 world-average (WA) value of $\tau(\Omega_c^0)=(69\pm12)$ fs extracted from fixed target experiments (for a short review of the $\Omega_c^0$ lifetime, see \cite{Cheng:2021vca}).
As a result, a new lifetime pattern emerged
\begin{equation} \label{eq:newhierarchy}
\tau(\Xi_c^+)>{\tau(\Omega_c^0)}>\tau(\Lambda_c^+)>\tau(\Xi_c^0).
\end{equation}

In 2019, LHCb has performed precision measurements of the $\Lambda_c^+$, $\Xi_c^+$ and $\Xi_c^0$ lifetimes \cite{LHCb:2019ldj} as displayed in Table \ref{tab:expt_lifetimes}. The $\Xi_c^0$ baryon lifetime is approximately 3.3 standard deviations larger than the 2018 WA value.
In 2021 LHCb \cite{LHCb:tauOmegac_2} reported a further new measurement using promptly produced $\Omega_c^0$ and $\Xi_c^0$ baryons with the results shown in the same table. Finally,  Belle II has disclosed in 2022  the new measurements of $\tau(\Lambda_c^+)$ \cite{Belle-II:Lambdac} and $\tau(\Omega_c^0)$ \cite{Belle-II:Omegac} and achieved the most precise result of the $\Lambda_c^+$ lifetime to date. The current WA values of charmed baryon lifetimes  in Table \ref{tab:expt_lifetimes} result from the averages of the data from LHCb (2021), PDG (2022) and Belle II (2022).

\begin{table*}[t]
\caption{Evolution of the charmed baryon lifetimes measured in units of ${\rm fs}$}
\label{tab:expt_lifetimes}
\begin{center}
\begin{tabular}{l c c c c} \hline \hline
 & $\tau(\Xi_c^+)$ & $\tau(\Lambda_c^+)$ & $\tau(\Xi_c^0)$ & $\tau(\Omega_c^0)$ \\
\hline
PDG (2004-2018) \cite{PDG2018}~~~ & ~$442\pm 26$ & $200\pm6$ & $112^{+13}_{-10}$ & $69\pm12$\\
LHCb (2018) \cite{LHCb:tauOmegac} & & & & ~~$268\pm26$~~ \\
LHCb (2019) \cite{LHCb:2019ldj} & $457\pm6$ & $203.5\pm2.2$ & $154.5\pm2.6$ & \\
PDG (2020) \cite{PDG2020} & ~~$456\pm5$~~ & ~~$202.4\pm3.1$~~ & ~~$153\pm 6$~~ & ~~$268\pm26$~~ \\
LHCb (2021) \cite{LHCb:tauOmegac_2} & & & $148.0\pm3.2$ & $276.5\pm14.1$ \\
PDG (2022) \cite{PDG2022} & ~~$453\pm5$~~ & ~~$201.5\pm2.7$~~ & ~~$151.9\pm 2.4$~~ & ~~$268\pm26$~~ \\
Belle II (2022) \cite{Belle-II:Lambdac,Belle-II:Omegac} & & ~~$203.20\pm1.18$~~ &  & ~~$243\pm49$~~ \\
WA values (2023) & ~~$453\pm 5$~~ & ~~$202.9\pm1.1$~~ & ~~$150.5\pm 1.9$~~ & ~~$273\pm12$~~ \\
\hline \hline
\end{tabular}
\end{center}
\end{table*}

On the theoretical side, lifetimes of the heavy baryons are commonly analyzed within the framework of heavy quark expansion (HQE), in which the inclusive rate of the heavy baryon $\B_Q$ is schematically represented by
 \begin{eqnarray} \label{eq:OPE}
 \Gamma({\cal B}_Q) = {G_F^2m_Q^5\over
192\pi^3}\left(A_0+{A_2\over m_Q^2}+{A_3\over
m_Q^3}+{A_4\over
m_Q^4}+\cdots\right),
 \end{eqnarray}
where
$G_F$ is the Fermi constant and $m_Q$ is the heavy quark mass. 
An analysis of the lifetimes of charmed baryons in HQE up to $1/m_c^4$ described by dimension-7 four-quark operators was  performed in  Ref. \cite{Charm:2018}. While the predictions on $\tau(\Lambda_c^+)$ and $\tau(\Xi_c^+)$ were improved after including corrections from dimension-7 operators, one of us (HYC) has introduced some additional parameters such as $y$ and $\alpha$ to the baryonic matrix elements to ensure the validity of HQE in the $\Omega_c^0$ sector. Although the value of $\tau(\Omega_c^0)\sim 2.3\times 10^{-13}s$ was predicted even before the first LHCb measurement of the $\Omega_c^0$ lifetime \cite{Cheng:HIEPA}, the use of the above-mentioned {\it ad hoc} parameters was not justified. In this work, we shall show that these difficulties can be circumvented. 

Through the optical theorem and HQE, the heavy hadron lifetimes are governed by the transition operators which have the following general expression
\begin{equation}\label{optical}
{\cal T} = \frac{G_F^2 m_Q^5}{192\pi^3}\left[ \left(
{\cal O}_3 + \frac{ 1}{m_Q^2}{\cal O}_5 + \frac{ 1}{m_Q^3}{\cal O}_6 \cdots 
\right)_2 +
\left(
\frac{ 1}{m_Q^3}
\tilde {\cal O}_6 + \frac{ 1}{m_Q^4} \tilde {\cal O}_7 \cdots 
\right)_4\cdots 
\right] \,,
\end{equation}
where
the subscript of the parenthesis denotes the number of quark operators.
A dimension-four operator is absent due the Luke's theorem~\cite{Luke:1990eg}.
The inclusive decay widths  evaluated by sandwiching the transition operator read as
\begin{equation}\label{width_1}
\Gamma({\cal B}_Q) = \frac{G_F^2m_Q^5}{192 \pi ^3 }  \left[
{\cal C}_3\left(  1 - 
\frac{\mu_\pi^2-\mu_G^2}{2 m_Q^2} \right) 
+ 2{\cal C}_5
\frac{  \mu_G^2}{m_Q^2}
+ {\cal C}_\rho\frac{ \rho_D^3}{m_Q^3} 
 \right]+  \Gamma_6 
 + \Gamma_7 \,.
\end{equation}
The Wilson coefficients
${\cal C}_3$, ${\cal C}_5$ and ${\cal C}_\rho$ depend only on the heavy flavor, while $\Gamma_{6,7}$ 
are proportional to $\langle \tilde {\cal O}_{6,7} \rangle$,
depending also on the light quarks in the heavy baryon. 
The quantities $\mu_\pi^2$, $\mu_G^2$ and $\rho_D^3$ are  of nonperturbative in nature, which will be introduced below.

Owing to the absence of lattice QCD calculations as input, one has to rely on the models to evaluate
the baryon matrix elements. Concerning the heavy baryon lifetimes, the non-relativistic quark model~(NRQM) turns out to be the favored one in the literature. 
Most of the baryon matrix elements can be extracted from the mass spectra, providing at least a certain way for the estimation. Nevertheless, the light quark masses are taken to be the constituent ones in $\langle {\cal O}_n \rangle$ but current ones in ${\cal C}_n$, yielding uncontrollable errors.

In this work, we adopt the bag model~(BM), where the up and down quarks are treated to be massless. The model did not receive much attention as  its estimate of the baryon matrix elements of four-quark operators  is much smaller than that of the NRQM.
Recently, we  found that once the unwanted center-of-mass motion~(CMM) of the bag model is removed in the heavy-flavor-conserving decays of heavy baryons, the four-quark operator matrix elements become twice larger~\cite{Cheng:2022jbr}.  In the meantime, there exist several issues for the NRQM estimates of the baryon matrix elements of two- and four-quark operators. In the spirit of HQET, two-quark operator matrix elements such as $\mu_\pi^2$, $\mu_G^2$ and $\rho_D^3$, and four-quark operator matrix elements such as $L^q_{{\cal B}_Q}$ to be defined below are also independent of the heavy quark mass $m_Q$. In the NRQM, $L_{{\cal B}_Q}^q$ is related to the heavy baryon wave function modulus squared at the origin; that is, $L_{{\cal B}_b}^q=-|\psi^{{\cal B}_b}_{b q}(0)|^2$.
Moreover, $|\psi^{{\cal B}_b}_{b q}(0)|^2\propto |\psi^B_{b\bar q}(0)|^2={1\over 12}f_B^2 m_B$ based
on the mass formula for hyperfine mass splittings of the bottom baryon and the $B$ meson.  
An immediate consequence is that the charmed baryon matrix element is much smaller than the bottom baryon one owing to the smallness of $f_D^2m_D$ compared to $f_B^2 m_B$. This feature is also manifested 
in the realistic NRQM calculation, see Table V below.  However, a large deviation between the values of $|\psi^{{\cal B}_c}_{c q}(0)|^2$  and $|\psi^{{\cal B}_b}_{b q}(0)|^2$ is not consistent with
the expectation of the heavy quark limit. This undesired feature can be overcome in the improved bag model which we are going to elaborate on in this work. 
Therefore, it is worthwhile to revisit the bag model for  the study of heavy baryon lifetimes.

This paper is organized as follows. In Sec.~\MakeUppercase{\romannumeral 2}, we use the bag model to compute the baryon matrix elements such as those of $\mu_\pi^2$, $\mu_G^2$ and $\rho_D^3$.
Special attention is paid to the running of the four-quark operator matrix elements. 
In Sec.~\MakeUppercase{\romannumeral 3}, we decompose  the  transition operator into several parts and briefly discuss  the Wilson coefficients therein.
Explicit expressions of the contributions to the decay widths of heavy baryons from four-quark operators at dimension-6 and -7 are given.
The numerical results are presented and discussed in Sec.~\MakeUppercase{\romannumeral 4} and we conclude this work in Sec.~\MakeUppercase{\romannumeral 5}.  Appendix A is devoted to the evaluation of four-quark operator matrix elements in the bag model.

	\section{Matrix elements in the  bag model}
	Under the heavy quark effective field theory~(HQET), the decomposition of the baryon 
	mass ~\cite{Neubert:1993mb}
	\begin{equation}\label{mugmupi}
		M_{{\cal B}_Q} = m _Q + \overline{\Lambda} 
		+\frac{\mu_\pi^2 }{2m_Q} - 
  C_G(m_Q, \mu) 
  \frac{\mu_G^2(\mu)}{2m_Q} + O(1/m_Q)\,,
	\end{equation}
	is related to the quantities 
	\begin{eqnarray}\label{massoperator}
		\mu_\pi^2 &\equiv& \langle \overline{Q}_v (i\vec{D})^2 Q_v \rangle_{{\cal B}_Q} =- \lambda_1   + O( 1/m_Q)\nonumber\\
		\mu_G^2 &\equiv&  \frac{g_s}{2}\langle \overline{Q}_v \sigma_{\mu\nu}  G^{\mu\nu} Q_v \rangle_{{\cal B}_Q} 
		=d_H{\lambda}_2 + O( 1/m_Q)\,,
	\end{eqnarray}
where $d_H=0, 4 $ for the antitriplet heavy baryon $T_Q$ and the sextet one $\Omega_Q$, respectively, $ig_sG_{\mu\nu} = [iD^\mu, iD^\nu]$ is the gluon field strength, $g_s^2 = 4\pi \alpha_s$, $\sigma_{\mu\nu} = \frac{i}{2}[\gamma^\mu,\gamma^\nu]$ and $Q = c,b$. 
We have taken the shorthand notation of $\langle {\cal O} \rangle _{{\cal B}_Q} \equiv 
	\langle {\cal B}_Q |  {\cal O} | {\cal B}_Q   \rangle / 2 M_{{\cal B}_Q }$  and it shall be noticed that the HQET parameters $\overline{\Lambda}$, $\lambda_1$ and ${\lambda}_2$ are independent of the heavy quark mass.
 The $Q_v$ is the heavy quark field defined as 
	\begin{equation}\label{Eq3}
		Q_v(x)
  + \frac{i\slash\hspace{-8pt}{D}_\perp}{2m_Q} Q_v  + O(1/m_Q^2)
  = e^{im_Q \vec{v}\cdot \vec{x}} Q(x)\,. 
	\end{equation}
	If not stated otherwise, 
 the matrix elements are evaluated at $x = 0 $ and  the first-order correction in Eq.~\eqref{Eq3} is treated as uncertainties, leading to $Q_v = Q$ when derivatives are not involved.
 Due to the reparametrization invariance~\cite{Abbott:1980hw}, $\mu_\pi^2$ does not get renormalized in the dimensional regularization. On the other hand, 
 $C_G(m_Q, \mu)$ is obtained by matching the full QCD theory with HQET  at the energy scale $\mu$, and 
 the renormalization dependence of $\mu_G^2$ is canceled by that of $C_G(m_Q,\mu)$~\cite{Grozin:1997ih}. 
The dependence of $m_Q$ in $C_G(m_Q,\mu)$ is generated by the renormalization effects known as  the Appelquist–Carazzone decoupling theorem~\cite{Appelquist:1974tg}. 
 
The mass corrections  $\overline{\Lambda} $, $\lambda_1$ and ${\lambda}_2$ correspond to the diquark energy,  kinetic energy of the heavy quark and chromomagnetic field energy, respectively. For $T_Q$,  as a working principle $\overline{\Lambda}$ is often taken to be equal among the baryons and $\lambda_2$ is extracted from the mass difference of $M_{\Sigma_Q} - M_{\Lambda_Q}$. Nevertheless, the diquarks in ${\Sigma_Q}$ and $\Lambda_Q$ are of spin-1 and spin-0 systems, respectively, and there is no reason to assume that their energies are the same. In general, the assumption of a universal $\overline{\Lambda}$ would cause errors  of  order $m_Q\overline{\Lambda}_\Delta/3$ to $\lambda_2$, where $\overline{\Lambda}_\Delta\approx 0.2$ GeV is the mass difference of spin-1 and spin-0 diquarks.
On the other hand, the assumption is acceptable in the $\Omega_Q$ sector, as the diquark system is of spin-1 for both $\Omega_Q$ and $\Omega_Q^*$.

In terms of the creation operators,
the baryon wave functions in the BM read 
\begin{eqnarray}\label{struc}
&&	|T_{Q}, \uparrow\rangle = \int\frac{1}{\sqrt{6} } \epsilon^{\alpha \beta \gamma} q _{a\alpha}^{\dagger} (\vec{x}_1) q_{b\beta}^{\prime \dagger}(\vec{x}_2) Q_{c\gamma}^\dagger (\vec{x}_3) \Psi_{A_\uparrow(qq'Q)}^{abc} (\vec{x}_1,\vec{x}_2,\vec{x}_3) [d^3  \vec{x}] | 0\rangle\,,\nonumber\\
&&	|\Omega_{Q}, \uparrow\rangle = \int\frac{1}{2\sqrt{3} } \epsilon^{\alpha \beta \gamma} s _{a\alpha}^{\dagger} (\vec{x}_1) s_{b\beta}^\dagger(\vec{x}_2) Q_{c\gamma}^\dagger (\vec{x}_3) \Psi_{S_\uparrow(ssQ)}^{abc} (\vec{x}_1,\vec{x}_2,\vec{x}_3) [d^3  \vec{x}] | 0\rangle\,,
\end{eqnarray}
where $(q,q') = (u,d), (u,s), (d,s)$ are the light-quark flavors of  $T_Q  \in \{ \Lambda_Q, \Xi_Q\} $, the Latin (Greek) alphabets stand for the spinor (color) indices, and $\Psi$ describes the spinor-spatial distributions of the quarks.
In the static bag limit, $\Psi$ is  given by 
\begin{eqnarray}\label{distri}
	&&\Psi_{A \uparrow (qq'Q)} ^{abc(SB)} (\vec{x}_1,\vec{x}_2,\vec{x}_3)= 
	\frac{{\cal N}}{\sqrt{2}}\Big( 
	\phi^a_{q\uparrow}(\vec{x}_1)\phi^b_{q'\downarrow}(\vec{x}_2)- \phi^a_{q\downarrow}(\vec{x}_1)\phi^b_{q'\uparrow}(\vec{x}_2)
	\Big) \phi^c_{Q\uparrow}(\vec{x}_3)
	\,,\nonumber\\
	&&\Psi_{S \uparrow (ssQ)} ^{abc(SB) } (\vec{x}_1,\vec{x}_2,\vec{x}_3)= 
	\frac{{\cal N}}{\sqrt{6}}\Big( 
	2 \phi^a_{s\uparrow}(\vec{x}_1)\phi^b_{s\uparrow}(\vec{x}_2) \phi^c_{Q\downarrow}(\vec{x}_3)\nonumber\\
	&&\quad\qquad- \phi^a_{s\downarrow}(\vec{x}_1)\phi^b_{s\uparrow}(\vec{x}_2) \phi^c_{Q\uparrow}(\vec{x}_3)- \phi^a_{s\uparrow}(\vec{x}_1)\phi^b_{s\downarrow}(\vec{x}_2) \phi^c_{Q\uparrow}(\vec{x}_3)
	\Big)
	\,,\nonumber\\
&& 
	\phi_{q\updownarrow}(\vec{x}) 
 = \left(
	\begin{array}{c}
u_q \chi _ \updownarrow\\
		iv_q  \hat{x} \cdot \vec{\sigma} \chi _ \updownarrow
	\end{array}
	\right)
\begin{cases}
 \left(
	\begin{array}{c}
		\omega_{q}^+j_0({\bf p}_q|\vec{x}|) \chi _ \updownarrow\\
		i\omega_{q}^- j_1 ({\bf p}_q|\vec{x}|) \hat{x} \cdot \vec{\sigma} \chi _ \updownarrow
	\end{array}
	\right),& \text{for}~|\vec{x}| < R,\\
    0,              & \text{otherwise},
\end{cases}
\end{eqnarray}
where ${\cal N}$ is a normalization constant, $\omega_q^\pm = \sqrt{E_q^k \pm m_q}$ with $E_q^k = \sqrt{{\bf p}_q^2 +  m_q^2} $, $j_{0,1}$ are the spherical Bessel functions, $\chi_\uparrow = (1,0)^T $, $\chi_\downarrow = (0,1)^T$ and $R$ is the bag radius.

In Eq.~\eqref{distri}, ${\bf p}_q$ is the magnitude of 
the 3-momentum of the bag quark $q$. From the  boundary condition, we have  \cite{DeGrand:1975cf}
\begin{equation}\label{7}
\tan \left({\bf p}_q R\right)=\frac{{\bf  p}_q R}{1-m_q R- E^k_q R}\,.
\end{equation}
The heavy quark limit can be obtained by taking the right-hand side of Eq.~\eqref{7} to be zero, leading to $\tan({\bf p}_Q R) = 0$ and 
${\bf p}_Q = \frac \pi R$.
To obtain the next-order correction, we plug ${\bf p}_Q = \frac \pi R (1+\frac{\xi} {m_QR}) + {\cal O}({1\over m_Q^2R^3}) $  into Eq.~\eqref{7} and obtain 
\begin{equation}
\tan\left(\pi + \frac{\xi}{m_Q R}\right) = -\frac{\pi}{2 m_Q R}  + {\cal O}\left(\frac{1}{m_Q^2R^2}\right)\,.
\end{equation}
Taking the Taylor expansion of the tangent function at $\pi$, we find 
\begin{equation}\label{pQ}
{\bf p}_Q = \frac{\pi }{R} \left( 1- \frac{1}{2m_QR} + {\cal O}\left({1\over m_Q^2R^2}\right)\right) \,. 
\end{equation}
The first-order corrections in charm and beauty baryons are $7\%$ and $2\%$, respectively, indicating that HQE works well for ${\bf p}_Q$.

\subsection{Two-quark operator matrix elements}
To the leading-order in $\alpha_s$,
the gluonic self-coupling does not contribute and hence gluon fields act as eight independent Abelian ones. 
The gluonic magnetic fields induced by $\phi_{q\lambda_q }$  are then given by~\cite{DeGrand:1975cf}
\begin{equation}\label{magnetic}
\vec{B}_{q\lambda_q}^a (\vec{x}) = \chi_{\lambda_q}^{ \dagger}\left(  \vec{\sigma}  \frac{\lambda^a}{4\pi}
\left(
2 M_q(r) + \frac{\mu_q(r)}{R^3} -\frac{\mu_q(r)}{r^3}  
\right)
+ \frac{3\lambda^a}{4\pi }\hat{r} ( \hat{r}\cdot \vec{\sigma}) \frac{\mu_q(r) }{ r^3 } \right) \chi_{\lambda_q} \,,
\end{equation}
where the definition of $M_q(r)$ and $\mu_q(r)$ can be found in Ref.~\cite{DeGrand:1975cf}, $\lambda^a$ are the Gell-mann matrices sandwiched by the quark color and $\lambda_q\in \{ \uparrow , \downarrow\} $. 
The interaction energy between $q_1$ and $q_2$ is equivalent to the energy 
stored in the magnetic field, given by~\cite{DeGrand:1975cf}
\begin{equation}\label{Magneticenergy}
\Delta E_M = -4\pi \alpha_s \sum_{\lambda_{1}, \lambda_{2}}
{\cal F}_S(\lambda_1, \lambda_2)
\sum_a \int d^3x \vec{B}^a_{q_1\lambda_{1}}(\vec{x}) \cdot \vec{B}^a_{q_2\lambda_{2}}(\vec{x})\,,
\end{equation}
where ${\cal F}_S(\lambda_{1},\lambda_{2})$ depends on the spin configurations. Explicitly, we have 
\begin{eqnarray}\label{sumoverspin}
&&\sum_{\lambda_{1}, \lambda_{2}}
{\cal F}_S (\lambda_1,\lambda_2)  (\chi^\dagger_{\lambda_{2}} \sigma_i  \chi_{\lambda_{2}} ) ( \chi^\dagger_{\lambda_{1}} \sigma_j  \chi_{\lambda_{1}}) =4  \vec{s}_1 \cdot{\vec{s}_2}  \,\delta_{ij}
\,,
\end{eqnarray}
where $\vec{s}_{1,2}$ are the spins of $q_1$ and $q_2$, and $4 \vec{s}_1\cdot\vec{s}_2 = -3 ,1$ for spin-0 and spin-1 configurations, respectively.
In the BM, $R^{-1}$ plays the role of a typical energy scale of the hadron, and the running of $\alpha_s$ with respect to $R$ is~\cite{Zhang:2021yul}
\begin{equation}\label{alphas}
\alpha_s = \frac{0.296}{\log(1+ (0.281R)^{-1})}\,. 
\end{equation}
In particular, we have $\alpha_s = (0.52, 0.55)$ for $R=(4.6, 5.0)\,{\rm GeV}^{-1}$. 
It is interesting to see that\footnote{The running of $\alpha_s(\mu)$ in this work is evaluated by using {\tt RunDec}~\cite{Herren:2017osy} with $(\mu_b,\mu_c) = (4.5$ GeV$,1.6$ GeV) and $\alpha_s(M_z)=0.1181$.
} $\alpha_s(1\,$GeV$)= 0.495$
is close to  Eq.~\eqref{alphas} with $R=5$~GeV$^{-1}$, which is a typical bag size of the baryon, indicating that the wave functions in the BM are applicable around the hadronic scale $\mu_H = 1$~GeV. 
In this work, we shall  allow $\mu_H$ to range from $0.8~$GeV to $1.2$~GeV to take into account the  hadronic uncertainties. 
By substituting Eqs.~\eqref{magnetic} and \eqref{sumoverspin} into Eq.~\eqref{Magneticenergy}, 
we find  
 $\overline{\Lambda}_\Delta  = 0.207$~GeV with $R=5$ GeV$^{-1}$ and $m_{q_1}=m_{q_2}=0$.

\begin{table}[t!]
	\caption{Inputs of the BM with $R$ and $m_q$ in units of   GeV$^{-1}$ and GeV, respectively \cite{Zhang:2021yul}. Here $m_q$ should not be confused with the pole quark mass}\label{para}
	\vspace{6pt}
	\begin{tabular}{cccccc||cccc}\hline 
		$R_{\Lambda_b}$ & $R_{\Xi_b}$ &$R_{\Omega_b}$ &$R_{\Lambda_c}$ & $R_{\Xi_c}$ &$R_{\Omega_c} $~~ &
		~~$m_{u,d}$&$m_s$ & $m_c$ & $m_b$  \\
		\hline
		$4.60$ &$ 4.71$ &$ 4.77$  & $4.86$ & $4.88$ & $4.93$~~ & ~~$0$ & $0.279$&$1.641$& $5.093$  \\
		\hline
	\end{tabular}
\end{table}

	\begin{table}[t!]
	\caption{The parameters $\lambda_1$ and $\lambda_2$ calculated in the bag model in units of GeV$^{2}$}\label{para2}
	\vspace{6pt}
	\begin{tabular}{c|cccccc}\hline 
&$\Lambda_b$& $\Xi_b$ &  $\Omega_b$ & $\Lambda_c$ & $\Xi_c$ & $\Omega_c$\\ 
		\hline
~$\lambda_1$~ &$-0.466$&$ -0.445$&$ -0.434$&$ -0.418$&$ -0.414$&$ -0.407$\\
~$\lambda_2$~ &0.0653&0.0585&0.0522&0.0553&0.0504&0.0450\\
		\hline
	\end{tabular}
\end{table}

The mass correction $\lambda_1$ corresponds to the kinetic energy of the heavy quark in the Newtonian limit with a minus sign. Therefore, we have 
\begin{equation}
\lambda_1 = - {\bf p}_Q^2 = -\frac{\pi^2}{R^2}\,. 
\end{equation}
 On the other hand,  $\lambda_2$ in Eq.~\eqref{mugmupi} describes the 
mass correction due to the 
 spin-spin interactions 
 between the heavy quark and the magnetic field. Comparing to Eq.~\eqref{Magneticenergy}, we  arrive at
 \footnote{
Since the right-hand side of Eq. (\ref{14}) is a  leading-order result in $\alpha_s$, we have taken $C_G(m_Q,\mu_H) = 1 $  on the left-hand side for reasons of consistency. 
Using integration by part and keeping the leading term in $1/m_Q$,
it is possible to show that Eq.~\eqref{massoperator} can be recasted to the form $\langle \vec{J}\cdot\vec{A} \rangle $, where $\vec{J} =g_s\overline{Q}\vec{\gamma} Q$ and $\vec{\nabla}\times \vec{A} = \vec{B}$ are the current density and vector potential, respectively. It is identical to $-g_s^2\langle \vec{B}\cdot \vec{B}\rangle $ to  the leading order in $\alpha_s$. }
 \begin{equation}\label{14}
 \frac{\lambda_2}{2 m_Q} = 4 \pi \alpha_s \sum _{a,q} \int d^3 x \vec{B}_{Q\uparrow}^a (\vec{x}) \cdot \vec{B}_{q\uparrow}^a (\vec{x})
 +O\left(\frac{1}{m_Q^2 R^2}\right)
 \,. 
 \end{equation}
 The  bag parameters are taken from Ref. \cite{Zhang:2021yul} and collected in Table~\ref{para}. The numerical results of $\lambda_{1,2}$ are listed in Table~\ref{para2}.
 It shall be noted that the final results depend weakly on the input of $m_Q$ in the BM. 
 Numerically, $\lambda_{1,2}$ vary less than 2\% if the pole quark masses given in Eq.~\eqref{kimas} below are used.
 In the limit of large $m_Q$, $\lambda_{1,2}$ shall be independent of the heavy quark mass $m_Q$, and therefore their values for ${\cal B}_b$ and ${\cal B}_c$ systems should be the same.  A small deviation of order $O(1/m_Q^2)$ comes from the fact that heavy baryons have different bag radii. As HQET is more reliable in ${\cal B}_b$, in the numerical evaluation we shall allow the values of ${\cal B}_c$ to vary from their original values to the ones of ${\cal B}_b$. For instance, we take $\lambda_1 = -(0.418\sim 0.466)$ GeV$^{2}$ for $\Lambda_c$, causing $10\%$ uncertainties. Additional $6\%$ and $14\%$ uncertainties from $1/m_Q^{2}$ are also included for ${\cal B}_b$ and ${\cal B}_c$, respectively, as described in Eq.~\eqref{pQ}. 
We note that the calculated $\lambda_{1}$ is consistent with the literature. On the contrary, $\lambda_2$ is about $40\%$ larger for $T_Q$, but the deviation will not reflect in the numerical results as $\lambda_2$ is always accompanied by $d_H$ in  Eq.~\eqref{massoperator} and $d_H = 0 $ for $T_Q$. 

Besides $\mu_\pi^2$ and $\mu_G^2$, there is an additional two-quark operator whose importance has been stressed recently, known as the Darwin operator 
\begin{eqnarray}\label{rhod}
\rho_D^3 &=&  {1\over 2M_{{\cal B}_Q}}\left\langle {\cal B}_Q|
\overline{Q}_v (iD_\mu ) ( i v\cdot D) (i D^\mu) Q_v |{\cal B}_Q\right\rangle\non\\
&=&
{ ig_s\over 2M_{{\cal B}_Q}}\left\langle {\cal B}_Q|
\overline{Q}_v (iD_\mu ) G^{0\mu} Q_v |{\cal B}_Q\right\rangle\,,
\end{eqnarray}
where we have used $v_\nu = (1,0,0,0)$ for the baryon at rest and the equation of motion $(i v\cdot D) Q_v = 0 $ in HQET.
In terms of the  matrix elements $L_{{\cal B}_Q} ^q$, $S_{{\cal B}_Q} ^q$ and $P_{{\cal B}_Q} ^q$ of the four-quark operators to be defined in the next subsection of Eq.~\eqref{4quark}, the Darwin operator
takes the form~\cite{Bigi:1993ex}\footnote{Considering the consistency with $\mu_G^2$, we take the running of $\alpha_s $ as in Eq.~\eqref{alphas}.}
\begin{equation}\label{rhod2}
\rho_D^3 
= -4\pi \alpha_s \sum_{q}\frac{1}{24} \left(
4L_{{\cal B}_Q} ^q  - \tilde{L}_{{\cal B}_Q} ^q  - 6 S_{{\cal B}_Q} ^q + 2 \tilde{S} _{{\cal B}_Q}  ^q  + 6 P_{{\cal B}_Q}^q - 2 \tilde{P} _{{\cal B}_Q}^q
\right) \,,
\end{equation}
where $q$ sums over the light quarks of ${\cal B}_Q$. In deriving the above equation, use of $4[D^\mu, G_{\mu\nu}]=-{g_s}\lambda^a\sum_q\overline{q}\lambda^a\gamma_\nu q$  has been employed.

By using Table \ref{scen} below in advance, the results of the two-quark operator matrix elements in the BM are summarized in Table~\ref{para3}, where the ones in the NRQM~\cite{Charm:2022,bottom:2023} are also included for comparison. 
We see that $\mu_\pi^2$, $\mu_G^2$ and $\rho_D^3$ all depend weakly on the heavy quark flavor in the BM, which is a good sign for HQET.  
The  values of $\mu_\pi^2 $ are compatible within the range of uncertainties, whereas 
the central value of $\mu_G^2$ for $\Omega_c$ deviates around 30$\%$.

\begin{table}[t!]
	\caption{The  two-quark operator matrix elements, where $\mu_{G,\pi}^2$ and $\rho_D^3$ are in units of $10^{-1}$GeV$^2$ and $10^{-2}$\,GeV$^3$, respectively. 
 The parameters of $\mu_G^2$ and $\rho_D^3$ are evaluated at the hadronic scale $\mu_H$ in the BM.
 Here, the numbers in the parentheses are the uncertainties counting backward in digits, for example, $4.42(24) = 4.42\pm 0.24$. The values  of the NRQM are quoted from Refs.~\cite{Charm:2022,bottom:2023}}\label{para3}
	\vspace{6pt}
	\begin{tabular}{lc|cccccc}\hline 
  {Model}
& &$\Lambda_b^0$& $\Xi_b^{0,-}$ &  $\Omega^-_b$ & $\Lambda^+_c$ & $\Xi_c^{0,+}$ & $\Omega_c^0$\\ 
		\hline&$\mu_\pi^2$~ &$4.66(28)$&$ 4.45(27)$&$ 4.34(80)$&$ 4.42 (81) $&$ 4.30 (80)$&$ 4.20 (80)$\\
BM~&$\mu_G^2 $~ &$0$& $0$ & $2.09(12)$&
$0$&
 $0$&
 $1.95(38) $\\
& $\rho_D^3$~ ~& ~2.29(23)&2.38(24)&2.66(27)&2.06(21)&2.22(22)&2.68(27)
 \\
		\hline
&$\mu_\pi^2$~ &$5.0(6)$&$ 5.4(6)$&$ 5.6(6) $&$ 5.0 (15) $&$ 5.5 (17)$&$ 5.5 (17)$\\
NRQM~&$\mu_G^2 $~ &$0$& $0$ & $1.93(68)$&
$0$&
 $0$&
 $2.6(8) $\\
& $\rho_D^3$~ & 3.1(9) &3.7(9) &5.0(21)&4(1)&5.5(20)&6(2)
 \\
 \hline
\end{tabular}
\end{table}

It is an appropriate place to discuss the renormalization-scale dependence of $\mu_G^2$. 
We note that the $m_Q$ dependence of $C(m_Q,\mu)$ can be factorized into two parts~\cite{Grozin:1997ih}
\begin{equation}
C_G (m_Q,\mu) = C_G (m_Q,m_Q) \exp \int_{\alpha_s(\mu)}^{\alpha_s(m_Q)} d \alpha_s \frac{\gamma\left(\alpha_s\right)}{\beta\left(\alpha_s\right)}\,, 
\end{equation}
where  the $\mu$ dependence is governed by $\gamma(\alpha_s)$ and $\beta(\alpha_s)$. 
In the BM,  both parameters and matrix elements are determined at the hadronic scale.
To employ the results for the lifetimes, we use
\begin{equation}\label{17}
C_G(m_Q,\mu_H) \mu_G^2(\mu_H)  = C_G (m_Q,m_Q)  \mu_G^2(m_Q) \,,
\end{equation}
derived from the renormalization-scheme independence of $C_G\mu_G$. 
The coefficient  $C_G(m_Q, m_Q)$ has been determined up to three-loop in Ref.~\cite{Grozin:2007fh} in the $\overline{\text{MS}}$ scheme, given by 
\begin{eqnarray}\label{18}
C_G (m_c,m_c) = 1.6506\,,~~~ C_G (m_b,m_b)=  1.2664\,. 
\end{eqnarray}
Although we can renormalize one of them as unity,  it cannot be done simultaneously for both in HQET. 
We note that $C_G (m_Q,m_Q)$ is missing in the previous study of the singly heavy baryon lifetimes and the formalism of $\mu_G^2$ shall be modified  as 
\begin{equation}
C_G (m_Q,m_Q)
\mu_G^2 (m_Q)
= \frac{2}{3}
\left(M_{\Omega_Q^*}^2 - M_{\Omega_Q}^2\right) \,. 
\end{equation}
The ratio of the mass splitting in $\Omega_Q$ is given by
\begin{equation}
R^G\equiv
\frac{ M_{{\Omega}_b^*}^2  - M_{{\Omega}_b}^2  }
{ M_{{\Omega}_c^*}^2  - M_{{\Omega}_c}^2  }= \frac{C_G(m_b,m_b)\mu_G^2(m_b) }{C_G(m_c,m_c )\mu_G^2(m_c)} = 0.6914\,, 
\end{equation}
where the anomalous dimension of the $\mu_G^2$ evolution can be found in Eq.~(13) of Ref.~\cite{Grozin:2007fh}. It is compatible with the experimental value\footnote{
We take $M_{\Omega_b^*}-M_{\Omega_b }=24.3$ GeV from Ref.~\cite{Karliner:2014gca}.} $R_{\text{exp}}^G=0.77$,  indicating that  the coefficient $C_G(m_Q,\mu)$ cannot be neglected.

The Darwin term $\rho_D^3$ in both the BM and NRQM obeys the same hierarchy $\Omega_Q > \Xi_Q > \Lambda_Q$, where the differences are induced by the strange-quark mass. More interestingly, $\rho_D^3$ shares similar values for $T_Q$ and $\Omega_Q$, which is anticipated from 
Eq.~\eqref{rhod} as only the heavy quark fields get involved\footnote{
Strictly speaking, in Eq.~\eqref{rhod} only the gluonic electric field of $G^{0\mu}$ participates in but not the magnetic one. 
As the gluonic electric (magnetic) field does not (does) depend on the light quark spins,   $T_Q$ and $\Omega_Q$ shall share a similar value of $\rho_D^3$ but not $\mu_G^2$. 
}. 
However, it is a non-trivial result in view of Eq.~\eqref{rhod2} as $L_{T_Q}^q$ and $L_{\Omega_Q}^q$ differ significantly.
It is interesting to point out that Refs.~\cite{Charm:2022} and \cite{bottom:2023} use $\alpha_s({\cal B}_c)=1$ and $\alpha_s ({\cal B}_b)= 0.22$ for $\rho_D^3$.  
However, the huge ratio $\alpha_s({\cal B}_c)/\alpha_s({\cal B}_b) = 4.5$ is compensated by the large difference in the four-quark operator matrix elements, resulting in a weak dependence of $\rho_D^3$ on $m_Q$. On the contrary,  both $\alpha_s$ and the four-quark operator matrix elements in the BM depend slightly on $m_Q$. The running of $\rho_D^3$  will be discussed in the next section.

Before ending this section, it has to be pointed out that 
the wave functions given in Eq.~\eqref{distri} encounter some inconsistencies as they are  localized in the three-dimensional space. According to the Heisenberg uncertainty principle, it cannot be a 3-momentum eigensate, which is referred to as the CMM problem. 
However, this problem can be neglected in $\lambda_{1,2}$.
As $\lambda_1$ is related to $\langle {\bf p}_Q^2\rangle$, it remains unchanged after removing CMM~\cite{Cheng:2022jbr}. The readers who are interested in the interplay between mass corrections and matrix elements are referred to Ref.~\cite{Zhang:2022bvl}. On the other hand, 
by an explicit calculation, we find that
the CMM will in general decrease $\lambda_2$ around 10\% for a fixed $\alpha_s$. Nonetheless, the effects can be compensated by increasing $\alpha_s$ by 10\%. As $\alpha_s$ is fitted from the mass spectra without removing CMM, it  should not be removed in the calculations of $\lambda_2$ either. 
For $\rho_D^3$, as we shall see shortly, the matrix elements are evaluated by removing the unwanted CMM. However, as $\alpha_s$ is likely to be underestimated by 10$\%$, we thus include $10\%$ uncertainties for $\rho_D^3$ in Table~\ref{para3}.

\subsection{Four-quark operator matrix elements}

Recall the shorthand notation for the matrix elements of an arbitrary operator  $\langle {\cal O}\rangle_{{\cal B}_Q} = \langle {\cal B}_Q| {\cal O} | {\cal B}_Q\rangle /2 M_{{\cal B}_Q}$.  
We parameterize the matrix elements of the dimension-6 four-quark operators as 
\begin{eqnarray}\label{4quark}
&&L_{{\cal B}_Q}^q \equiv \left \langle \left({Q}_\alpha^\dagger L^\mu q_\alpha \right)\left({q}^\dagger _\beta  L_\mu Q_\beta \right)\right \rangle_{{\cal B}_Q} \,,\qquad \tilde{L}_{{\cal B}_Q}^q \equiv \left \langle \left({Q}_\alpha^\dagger L^\mu q_\beta \right)\left({q}^\dagger _\beta  L_\mu Q_\alpha \right) \right \rangle_{{\cal B}_Q}\,,  \nonumber    \\
&&S_{{\cal B}_Q}^q \equiv \left \langle \left(\overline{Q}_\alpha q_\alpha \right)\left(\overline{q}_\beta Q_\beta \right)\right \rangle_{{\cal B}_Q} \,,\qquad \qquad~~~ \tilde{S}_{{\cal B}_Q}^q \equiv \left \langle \left(\overline{Q}_\alpha q_\beta  \right)\left(\overline{q}_\beta Q_\alpha \right)\right \rangle_{{\cal B}_Q}\,,      \\
&& P_{{\cal B}_Q}^q \equiv \left \langle \left(\overline{Q}_\alpha \gamma_5 q_\alpha \right)\left(\overline{q}_\beta\gamma_5  Q_\beta \right)\right \rangle_{{\cal B}_Q} \,,\qquad ~~~ \tilde{P}_{{\cal B}_Q}^q \equiv \left \langle \left(\overline{Q}_\alpha \gamma_5  q_\beta  \right)\left(\overline{q}_\beta \gamma_5  Q_\alpha \right)\right \rangle_{{\cal B}_Q}\,,   \nonumber
\end{eqnarray}
and
\begin{eqnarray}
\tilde {I}_{{\cal B}_Q} ^q =-\tilde{B} {I}_{{\cal B}_Q} ^q \,,
\end{eqnarray}
where $L^\mu = \gamma^0 \gamma^\mu(1-\gamma_5)$ and $I_{{\cal B}_Q} ^q \in \{ L_{{\cal B}_Q} ^q,S_{{\cal B}_Q} ^q,P_{{\cal B}_Q} ^q\}$.
As strong interactions conserve parity, we do not need to consider the parity-violating operators, {\it i.e.}  $\langle \overline{Q} \gamma_5 q \overline{q}Q\rangle=0$. 
Although not written explicitly, it is understood that
$I_{{\cal B}_Q}^q$ depend on the energy scale. Our strategy is to first evaluate them at  the hadronic scale $\mu_H$, where the valence quark approximation of $\tilde B=1$ is automatically satisfied by the color structure in Eq.~\eqref{struc}.  After that, we evolve them to the heavy quark scale $\mu_Q$ to compute the lifetimes.

On the other hand, for the dimension-7 four-quark operators \cite{Lenz:2013aua}
\begin{equation}
\begin{aligned}
&P_1^q= m_q\,\overline{Q} (1-\gamma_5)q \overline{q}(1-\gamma_5)Q \,, & P_2^q = 
m_q\,\overline{Q}  (1+\gamma_5)q \overline{q}(1+\gamma_5)Q \,,~~~~~~~~ \\
&P_3^q= \frac{1}{m_Q}  {Q}^\dagger  \overleftarrow{D}_\rho L_\mu D^\rho q {q}^\dagger L^\mu  Q\,, & \quad P_4^q = \frac{1}{m_Q} \overline{Q} \overleftarrow{D}_\rho  (1-\gamma_5)D^\rho q \overline{q}  (1+\gamma_5) Q \,,
\end{aligned}
\end{equation}
and $\tilde P_i^q$ $(i=1,\cdots,4)$ obtained from $P_i^q$ by interchanging the colors of $\overline{q}$ and $Q$,
their matrix elements can be expressed in terms of that of dimension-6 ones: 
\footnote{The operators $P_1^q$ and $P_2^q$ have the same baryonic matrix elements as the latter is related to the former by hermitian conjugation; that is, $P_2^q=(P_1^q)^\dagger$
\cite{King:2021xqp}.}
\begin{eqnarray}\label{67relations}
&&\langle
	P_{i}^q \rangle_{ {\cal B}_Q}   = m_q \left(
	S_{{\cal B}_Q} ^q + P_{{\cal B}_Q} ^q 
	\right) ~~\text{for}~i = 1,2\,,\nonumber\\ 
	&& \langle  
	P_{3}^q \rangle_{ {\cal B}_Q} =  E_q L_{{\cal B}_Q} ^q \,,~~~\langle 
	P_{4}^q \rangle_{ {\cal B}_Q}  = E_q (S_{{\cal B}_Q} ^q  - P_{{\cal B}_Q} ^q ) \,,
\end{eqnarray}
and
\begin{equation}
	\langle\tilde P_{i}^q \rangle_{ {\cal B}_Q}  = -\tilde \beta_i \langle P_{i}^q \rangle_{ {\cal B}_Q}  ~~\text{for}~i = 1,\cdots,4\,.
\end{equation}
 where $\tilde{\beta}_i=1$ under the valence quark approximation. 
	In deriving Eq.~\eqref{67relations}, we have applied the approximation 
	\begin{eqnarray}\label{approx}
		&&\frac{1}{m_Q}\left \langle \left( \overline{Q} \overleftarrow{D}_\rho \Gamma D^\rho q\right)  \overline{q} \Gamma  Q\right \rangle =
		\frac{1}{m_Q} \left \langle \left( \overline{Q} \overleftarrow{\partial }_\rho  \Gamma \partial^ \rho  q \right) \overline{q} \Gamma  Q\right \rangle  + O(\alpha_s ) \nonumber\\
		&&=\frac{1}{m_Q} \left \langle \left( \overline{Q} \overleftarrow{\partial }_t  \Gamma \partial_t  q \right) \overline{q} \Gamma  Q\right \rangle+ O(\alpha_s ) + O(1/m_Q)\approx E_q \left \langle  \overline{Q}   \Gamma   q  \overline{q} \Gamma  Q\right \rangle \,,
	\end{eqnarray}
	where  $E_q$ is the energy of the bag quark, taken to be  $E_{u,d}  = M_p/3 \approx 0.32$~GeV and $E_{s}= E_{u,d} + (M_{{\Xi}_Q} - M_{\Lambda_Q}) =0.50$ GeV.
The spatial derivatives have been omitted as they are proportional to $O(1/m_Q)$ and  the last equation follows as the bag quarks are approximately in the energy eigenstates with $E_Q=m_Q + O(1/ m_Q)$.\footnote{Strictly speaking, $E_q$ can only be defined when  quark energies are not entangled, {\it i.e.} interactions between quarks are negligible.
In the BM, it is true by constructions in Eq.~\eqref{distri} where quark wave functions are independent to each other. In addition, the 
normalization condition in Eq.~\eqref{normalization} requires that $E_u + E_d + m_c = M_{\Lambda_c}$ which is satisfied with the use of $m_c$ in Table~\ref{para}. 
}
We note that similar approximations have also been employed in Ref.~\cite{Charm:2022} by substituting $\Lambda_{\text{QCD}}=0.33$~GeV for $E_q$.\footnote{The relation $\left\langle P_3^q\right\rangle\approx (p_Q\cdot p_q/m_Q) \left \langle  \overline{Q}   \Gamma   q  \overline{q} \Gamma  Q\right \rangle$ has been used in Refs. \cite{Charm:2018,Charm:2022} with $p_Q\cdot p_q$  taken to be $\sim m_Q\Lambda_{\rm QCD}$ in Ref. \cite{Charm:2022} and ${1\over 4}m_Q[(m_{{\cal B}_Q}^2-m_{\rm diq}^2)/m_Q^2 -1]$ in Ref. \cite{Charm:2018} with $m_{ \rm diq}$ being the diquark mass.}  The choice of $E_{u,d} = 0.32$~GeV was proved to be suitable in describing the exclusive semileptonic decays of heavy baryons~\cite{Zhang:2022bvl,Geng:2022fsr}.

As mentioned in the previous section, the wave functions in Eq.~\eqref{distri} are problematic as they are localized, inducing a nonzero CMM. To compute the four-quark operator matrix elements, we have to remove the CMM for a consistent procedure.
To this end, we have to distribute the wave functions homogeneously over all the space. Consequently, Eq.~\eqref{distri} is modified as 
 \begin{equation}
\Psi^{(HB)}(\vec{x}_1, \vec{x}_2 , \vec{x}_3) 
=
\int d^3  \vec{x}_\Delta
\Psi^{(SB)} (\vec{x}_1 - \vec{x}_\Delta, \vec{x}_2- \vec{x}_\Delta  , \vec{x}_3 - \vec{x}_\Delta )\,. 
 \end{equation}
Here, $(HB)$  and 
$(SB)$ stand for the homogeneous bag and static bag approaches, respectively. 
It is straightforward to show that the wave function is invariant under the space translation
\begin{equation}
\Psi^{(HB)}( \vec{x}_1 + \vec{d} , \vec{x}_2+ \vec{d}, \vec{x}_3+ \vec{d} )
=\Psi^{(HB)}( \vec{x}_1  , \vec{x}_2 , \vec{x}_3 )\,,
\end{equation}
which is not respected by $\Psi^{(SB)}$. 
A great advantage of $\Psi^{(HB)}$ is that it does not require additional parameter input. However, the calculation  becomes much more tedious. The methodology of the computation is given in Appendix A.

 \begin{table}[t!]
	\caption{ The matrix elements of the four-quark operators in units of $10^{-3}$~GeV$^{3}$ with $q_I=u,d$ evaluated at the hadronic scale $\mu_H$. 
			The results of the NRQM 	are quoted from Refs.~\cite{Charm:2022} and \cite{bottom:2023} for charm and beauty baryons, respectively}\label{scen}
		\vspace{6pt}
		\begin{tabular}{lc|rrrr|rrrr}\hline 
			{Model}~~& {~$({\cal B}_Q, q)$}~ & ~$(\Lambda_b, q_I )$~ & $ (\Xi_b,q_I)$&$ (\Xi_b,s)$ & ~$(\Omega_b,s)$~ & ~$(\Lambda_c, q_I )$ & $ (\Xi_c,q_I)$&$ (\Xi_c,s)$ & $ (\Omega_c,s)$ \\
			\hline
			\multirow{3}{*}{BM\,\footnote{
~Corresponding to the results in which the CMM is removed from the bag model. 
   }}			&$L_{{\cal B}_Q}^q$&$-5.44$&$-5.15$&$-5.88$ & $-34.12$~ & $-4.83$&$-4.87$&$-5.34$&$-31.63$\\
		&$S_{{\cal B}_Q}^q$ &$2.44$&$2.32$&$2.74$& $-5.41$~ &$1.96$&$1.98$&$2.32$&$-4.65$\\
			&$P_{{\cal B}_Q}^q$& $-0.27$&$-0.25$&$ -0.20$&$-0.62$~ & $-0.44$&$-0.44$&$-0.34$&$-1.12$\\
			\hline
			\hline
			\multirow{3}{*}{NRQM}
			&$L_{{\cal B}_Q}^q$&$ -13(5) $&$ -14(5) $&$ -18(6) $&$ -126(60) $~ &$ -5.1(15) $&$ -5.4(16) $&$ -7.4(22) $&$ -46(14) $\\
         &$S_{{\cal B}_Q}^q$ &$ 7(2) $&$ 7(2) $&$ 9(3) $&$ -21(10) $~ &$ 2.5(8) $&$ 2.7(8) $&          $3.7(11)$  &  $ -7.7(23) $\\
			&$P_{{\cal B}_Q}^q$&0&0&0& 0~ & $0$& $0$& $0$& $0$\\
			\hline
		\end{tabular}
	\end{table}	

 The results of $I_{{\cal B}_Q}^q$ in the BM and NRQM~\cite{bottom:2023,Charm:2022} are summarized in Table~\ref{scen}. 
As stressed in passing, $I_{{\cal B}_Q}^q$ depends on the energy scale and the results exhibited in the table are evaluated at $\mu_H$, where the valence quark approximation is valid. 
Besides the uncertainties from the parameter input,
additional 30$\%$ uncertainties
are put by hand in the NRQM to be conservative as described in Refs.~\cite{bottom:2023, Charm:2022}. 
Several remarks are in order:
\begin{itemize}
\item The $SU(3)$ flavor symmetry breaking can be examined by comparing $I_{\Xi_Q}^{q_I}$ and $I_{\Xi_Q}^{s}$. In the BM and NRQM, the breaking effects are around $10\%$ and $30\%$, respectively. 
\item 
Since the light quark $q$ has to be left-handed due to chiral symmetry, it is  the linear combination of $S_{{\cal B}_Q}^q - P_{{\cal B}_Q}^q  $ rather than $S_{{\cal B}_Q}^q + P_{{\cal B}_Q}^q  $ that appears in  the lifetimes to the leading-order of four-quark operators. Therefore, it suffices to consider $S_{{\cal B}_Q}^q - P_{{\cal B}_Q}^q  $ in discussing the finite heavy quark mass corrections. From the table, we see that $L_{{\cal B}_Q}^q$ and $S_{{\cal B}_Q}^q - P_{{\cal B}_Q}^q  $ in the BM vary less than 10$\%$ with respect to the heavy flavor. Accordingly, we  shall assign 10$\%$  uncertainties to $I_{{\cal B}_b}^q$  when computing the lifetimes. 
\item In the NRQM, $L_{{\cal B}_Q}^q$ is related to the heavy baryon wave function modulus squared at the origin, for example, $L_{\Lambda_b}^q=-|\psi^{\Lambda_b}_{b q}(0)|^2$~\cite{Blok:1991st}. From Table \ref{scen} we see that $|\psi^{\Lambda_b}_{b q}(0)|^2$ is of order $1.3\times 10^{-2}\,{\rm GeV}^3$ for the bottom baryon $\Lambda_b^0$, while
 $|\psi^{\Lambda_c}_{c q}(0)|^2=0.51\times 10^{-2}\,{\rm GeV}^3$ for $\Lambda_c^+$.
This is very annoying as $|\psi(0)|^2$ for hyperons is of the same order of magnitude as the bottom 
baryons (see Ref. \cite{LaYaou} for detail). Thus it is not comfortable to have the charmed baryon wave function at the origin substantially smaller than those in bottom and hyperon systems.\footnote{
In order to circumvent such inconsistency in the NRQM,   one of us (HYC) was forced to introduce an additional parameter $y=7/4$ by hand in Ref.~\cite{Charm:2018} to enhance $|\psi(0)|^2$ for charmed baryons as the large $m_Q$ limit is expected to work better for the beauty quark. 
Unfortunately, this will lead to a negative semileptonic decay width for $\Omega_c$, and an additional free parameter $\alpha$ was introduced in Ref.~\cite{Charm:2018} to rescue the inconsistency. In this work, we find that there is no need to introduce both {\it ad hoc} parameters $\alpha$ and $y$.  
}
Fortunately, this is no longer an issue in the BM where $L_{{\cal B}_Q}^q$ varies less than 10\% from the bottom to the charm sector.

\item Both the BM and NRQM agree with each other in $I_{{\cal B}_c}^q$ but differ largely in $I_{{\cal B}_b}^q$. 
Meanwhile, the HQET  and QCD sum rules give 
smaller values of 
${ L}_{\Lambda_b}^q = 
-(3.2\pm 1.6)$ ~\cite{Colangelo:1996ta} and $- (2.38
\pm 0.11 \pm 0.34\pm 0.22)$ ~\cite{Zhao:2021lzd}, respectively, in units of $10^{-3}$~GeV$^{3}$, which are much closer to the values of BM than that of NRQM.  
\end{itemize}

To compute the lifetimes, we have to evolve $I_{{\cal B}_Q}^q$ to the energy scale $\mu_Q$ of the heavy quark where the formalism is derived. 
Unfortunately, the matrix elements of the four-quark operators diverge to the leading-order of $\alpha_s$. To make sense out of the calculation, one has to regularize them. In turn, subtracting the infinity induces a renormalization-scheme dependence.
In HQET, the heavy quark $Q$ is treated as a static one. 
The renormalization-group evolution of the four-quark operator matrix elements are then given by~\cite{Neubert:1996we}
\begin{eqnarray}\label{45}
&&I_{{\cal B}_Q}^q (\mu_Q) = \kappa
I_{{\cal B}_Q}^q   (\mu_H) 
- \frac{1}{3} (\kappa-1) \tilde{I} _{{\cal B}_Q}^q (\mu_H)\,, \nonumber\\
&& \tilde{I} _{{\cal B}_Q}^q (\mu_Q) =  \tilde{I} _{{\cal B}_Q}^q (\mu_H) \,,
\end{eqnarray}
where $\kappa = \sqrt{ \alpha_s(\mu_H) /\alpha_s(\mu_Q) }$ and the relation
$2 t_{ij}^at_{kl}^a =\delta_{il}\delta_{jk} - \delta_{ij}\delta_{kl}/3$ has been used in the derivation. 
We then arrive at
\begin{eqnarray}\label{46}
I_{{\cal B}_b}^q(\mu_b) = (1.56 \pm 0.19) I_{{\cal B}_b}^q(\mu_H)\,,~~~
I_{{\cal B}_c}^q(\mu_c) = (1.26\pm 0.15) I_{{\cal B}_c}^q(\mu_H)\,,
\end{eqnarray}
whereas $\tilde{I}_{{\cal B}_Q}^q$ remains unchanged, leading to $\tilde B (\mu_Q)= 0.64 \pm 0.09$ and $0.79\pm 0.11$ for ${\cal B}_b$ and ${\cal B}_c$, respectively. 

 To explore the renormalization-scheme dependence of $I_{{\cal B}_Q}^ q$, we consider the 
full QCD operator matrix elements 
in the $\overline{\text{MS} }$ scheme.
The evolution of $I_{{\cal B}_Q}^q$ can be obtained by the fact that the anomalous dimension matrix is diagonalized in 
the operator basis $Q_\pm = (Q_2\pm Q_1 )/2 $ with~\cite{Buras:2020xsm}
\begin{equation}
Q_1 = 
\left({Q}_\alpha^\dagger L^\mu q_\alpha  \right)\left({q}^\dagger _\beta  L_\mu Q_\beta  \right)\,,
\qquad
Q_2 = 
\left({Q}_\alpha^\dagger L^\mu q_\beta \right)\left({q}^\dagger _\beta  L_\mu Q_\alpha \right)\,.
\end{equation}
Therefore, we shall have 
\begin{equation}
C_\pm ( \mu_1 ) \langle Q_\pm \rangle (\mu_1) = C_\pm ( \mu_2 ) \langle Q_\pm \rangle (\mu_2) \,,
\end{equation}
where $C_\pm ( \mu) $ are the Wilson coefficients and we have neglected the mixing with the penguin operators. The equality is derived by the fact that the amplitudes shall not depend on $\mu$. 
Taking $(\mu_1,\mu_2) = (\mu_Q,\mu_H)$, we obtain
\begin{equation}\label{49}
C_\pm ( \mu_Q ) \left( \tilde{L}_{{\cal B}_b}^q(\mu_Q) \pm 
{L}_{{\cal B}_b}^q(\mu_Q)
\right) = C_\pm ( \mu_H )\left( \tilde{L}_{{\cal B}_b}^q(\mu_H) \pm 
{L}_{{\cal B}_b}^q(\mu_H)
\right)  \,.
\end{equation}
In this renormalization scheme, a straightforward conclusion is that
$\tilde{L}_{{\cal B}_b}^q(\mu_H) + 
{L}_{{\cal B}_b}^q(\mu_H) = 0
$ leads to  $\tilde{L}_{{\cal B}_b}^q(\mu_Q) +
{L}_{{\cal B}_b}^q(\mu_Q)=0$ so long as $C_+ \neq 0 $. In other words, $\tilde{B}(\mu) = 1$ holds independently of $\mu$ for $L_{{\cal B}_Q}^q$ in this scheme. Let us return back to Eq.~\eqref{49} and obtain
\begin{equation}\label{50}
L_{{\cal B}_Q}^q(\mu_Q)  = 
\frac{C_- (\mu_H )}{C_- (\mu_Q )}
L_{{\cal B}_Q}^q(\mu_H) = 
U_- (   \mu_H, \mu_Q) 
L_{{\cal B}_Q}^q(\mu_H) 
\,.
\end{equation}
 In the leading-logarithmic approximation, the results are 
 \begin{eqnarray}\label{51}
&&U_-(\mu_H,\mu_b) = 
\left[
\frac{\alpha_s(\mu_H)}{\alpha_s(\mu_c)}
\right]^{\frac{4}{9}} 
\left[
\frac{\alpha_s(\mu_c)}{\alpha_s(\mu_b)}
\right]^{\frac{12}{25}} = 1.51\pm 0.16
\,,
\nonumber\\
&&U_-(\mu_H,\mu_c) = 
\left[
\frac{\alpha_s(\mu_H)}{\alpha_s(\mu_c)}
\right]^{\frac{4}{9}} = 1.22 \pm 0.13 \,.
 \end{eqnarray}
We see that the evolution of $L_{{\cal B}_Q}^q$ are similar in both  schemes, where  $L_{{\cal B}_b}^q$ and  $L_{{\cal B}_c}^q$ are enhanced by $50\%$ and $20\%$, respectively. 
Accordingly, we assume the evolution in Eq.~\eqref{50} is also applicable for both $S_{{\cal B}_Q}^q$ and $P_{{\cal B}_Q}^q$. 


\section{Decay widths}
The nonperturbative baryonic matrix elements are collected in Tables~\ref{para3} and \ref{scen}   and their values at the energy scale $\mu_Q$ are obtained through Eqs.~\eqref{17}, \eqref{45} for HQET and  \eqref{50} for full QCD. 
With these building blocks, we are  in a position to compute the lifetimes of heavy baryons.  In the beauty baryon decays, $(m_s/m_b)^2$ can be safely neglected, while $x_b=(m_c/m_b)^2$ is kept. Likewise, $x_c = (m_s/m_c)^2 $ shall be taken into account in charmed baryon decays. 
We briefly discuss the expressions of  ${\cal C}_{n}$, $\tilde {\cal O}_6$ and $\tilde {\cal O}_7$ with $n=3,5, \rho$ 
from the literature.

\subsection{Two-quark operators}

\begin{figure}[t]
\begin{center}
\includegraphics[width=0.28\linewidth]{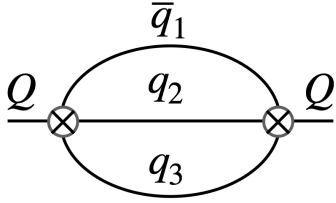}
\caption{The leading-order Feynman diagram for ${\cal C}_3$, where the $\otimes$ represents the insertion of the effective Hamiltonian for $Q\to \overline q_1{q}_2 q_3$}
\label{fig:c3}
\end{center}
\end{figure}

 As the experiments are able to probe the semileptonic inclusive decay widths, it is natural to decompose ${\cal C}_n$ into the form 
\begin{equation}
{\cal C}_n={\cal K}^{\text{SL}} _n+  \sum_{i,j=1 }^6 c_ic_j{\cal K}_{n,ij}^{\text{NL}} 
\,,
\end{equation}
where $c_1$ and $c_2$ are the leading Wilson coefficients for the effective Hamiltonian
\begin{equation}
{\cal H}_{eff} = \frac{G_F}{\sqrt{2}}\xi_Q\left[
c_1 (q_2^\dagger L^{\mu} q_1) ( q^\dagger_3L_\mu Q) + c_2
(q_3^\dagger L^{\mu} q_1) ( q^\dagger_2L_\mu Q) 
\right]+ O(\alpha_s) \,,
\end{equation}
with $\xi_Q$ being the CKM matrix elements, and 
 $c_{3\text{-}6}$ are the Wilson coefficients for QCD penguin operators. Note that the byproducts of $c_ic_j$ with $i,j\ge 3$ are discarded in the 
 next-to-the-leading order~(NLO). To evaluate ${\cal C}_3$, we take 
$Q\to \overline q_1 q_2 q_3$
as an illustration, depicted in Fig.~\ref{fig:c3}, where the $\otimes$ represents the insertion of ${\cal H}_{eff}$. 
To the leading-order~(LO), ${\cal C}_3$ can be obtained by matching Fig.~\ref{fig:c3} to ${\cal C}_3\overline{Q}Q$. Clearly, the coefficient ${\cal C}_3$ depends on the masses of $q_{1,2,3}$ appearing in the loop integral.
To the NLO with $i,j \in \{ 1,2\}$,
the results of ${\cal K}_3$ with a single massive $q_3$ can be found in Ref.~\cite{Bagan:1994zd}, and the doubly massive cases with $q_1=q_3$ are available in Ref.~\cite{Bagan:1994qw}. 
The expressions with semitauonic decays at the LO are found in Refs.~\cite{Mannel,Falk:1994}.
The coefficients for the penguin operators are given in 
Ref.~\cite{Krinner:2013cja}. 
On the other hand, ${\cal C}_5$ is available at NLO for the semileptonic  and  nonleptonic decays with massive and massless final state quarks, respectively~\cite{Mannel:2015jka,Mannel:2023zei}.
We  use the LO coefficients
with massive $q_1$ and $q_3$~\cite{Bigi:1992su,Blok:1992hw}. We note that
${\cal C}_5$ slightly affects the inclusive decay  widths. 
In particular, $\mu_G^2=0$  for $T_Q$ so their results are not affected by ${\cal C}_5$.
On the other hand, the explicit values of ${\cal K}_{D,ij}^{\text{NL}}$ and  
${\cal K}_D^{\text{SL}}$ are given in Refs.~\cite{SEMI_Bottom, NON_Bottom,Lenz:2020oce,King:2021xqp}, where we shall only use the LO results.

\subsection{Four-quark operators}
\begin{figure}[t]
\begin{center}
  \begin{subfigure}{0.3 \linewidth}
 \includegraphics[width=\linewidth]{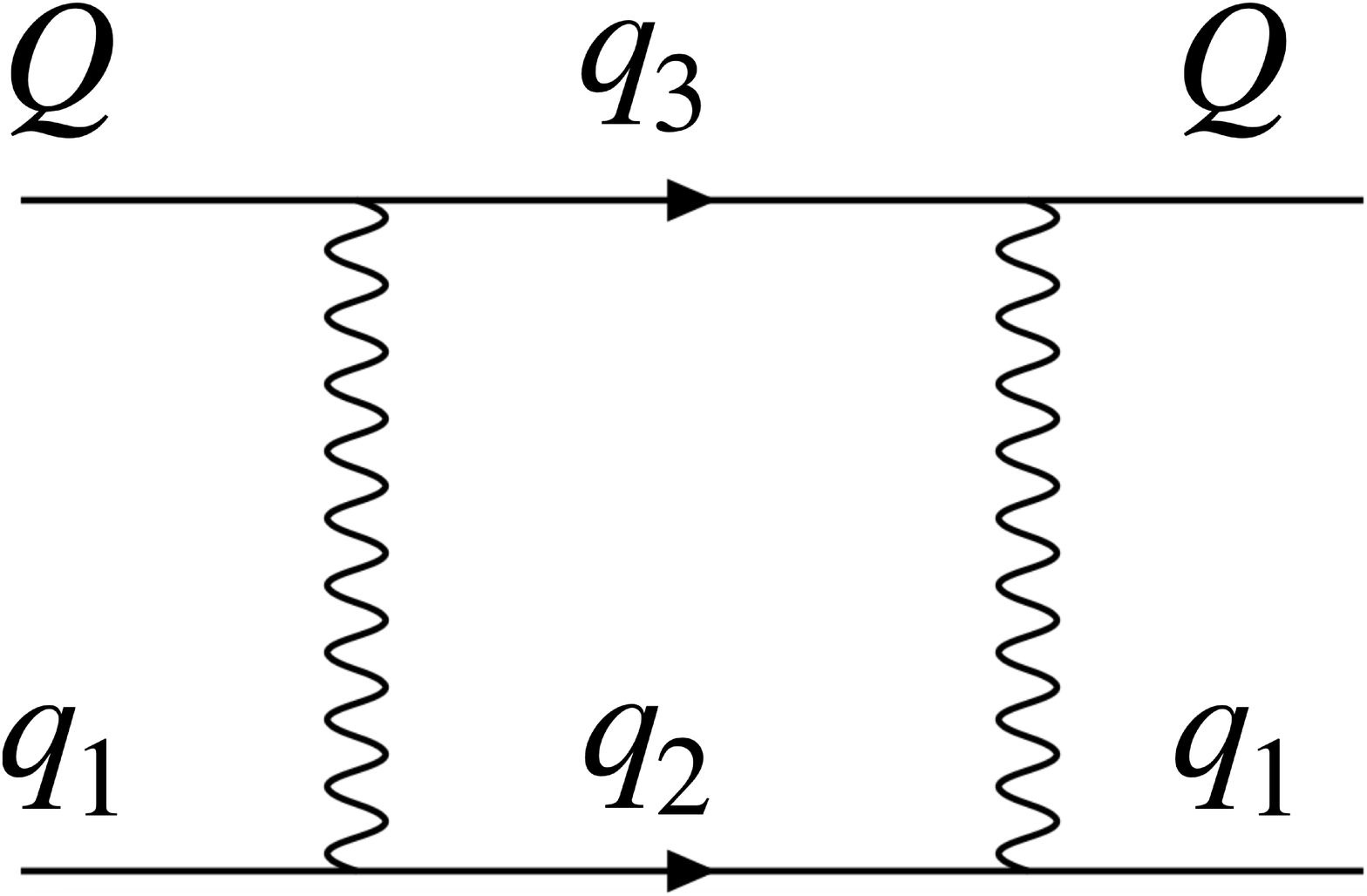}
    \caption{} \label{fig:2a}
  \end{subfigure}~~~
  \begin{subfigure}{0.3 \linewidth}
 \includegraphics[width=\linewidth]{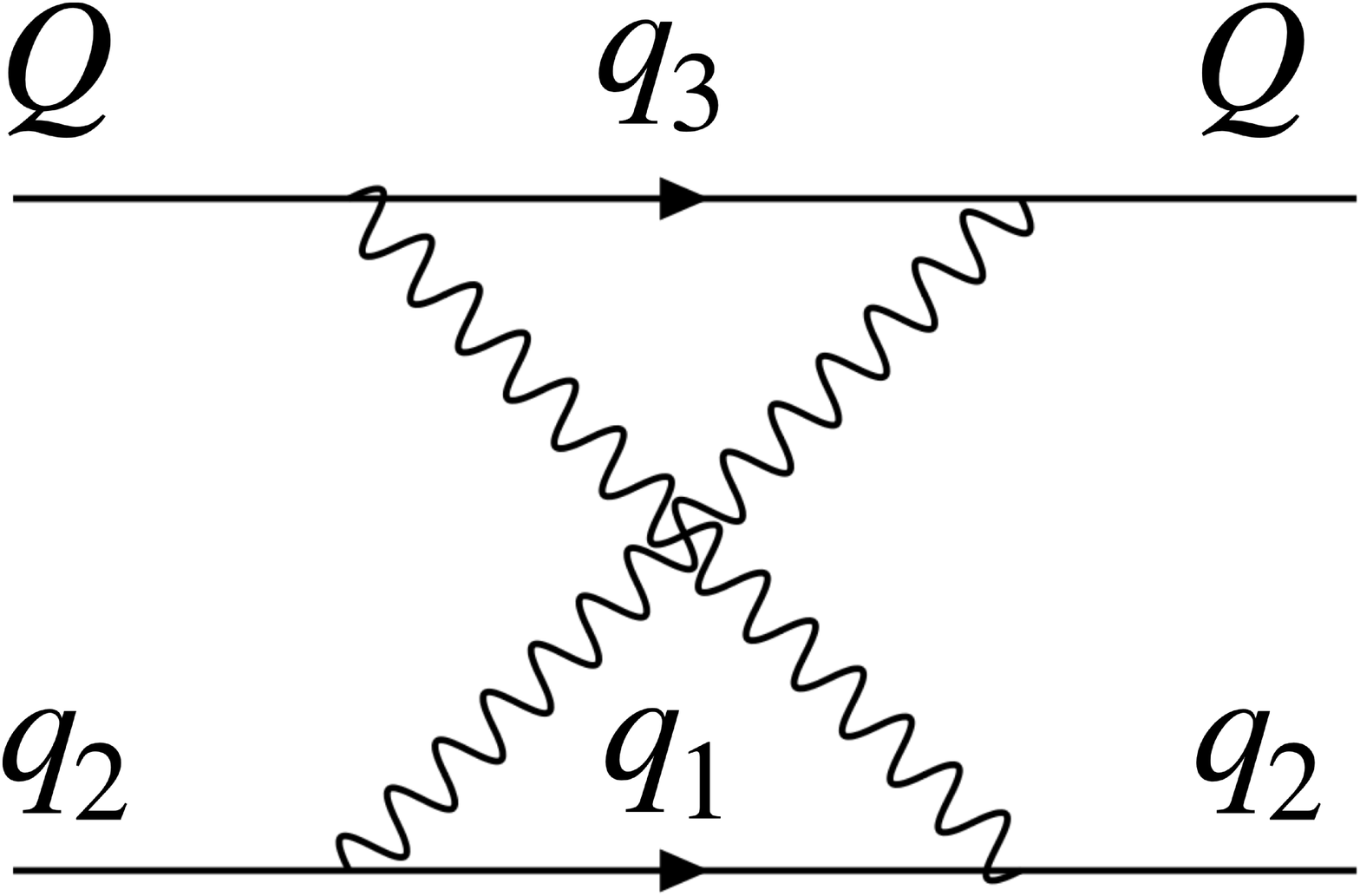}
    \caption{} \label{fig:2b}
  \end{subfigure}~~~
  \begin{subfigure}{0.3 \linewidth}
 \includegraphics[width=\linewidth]{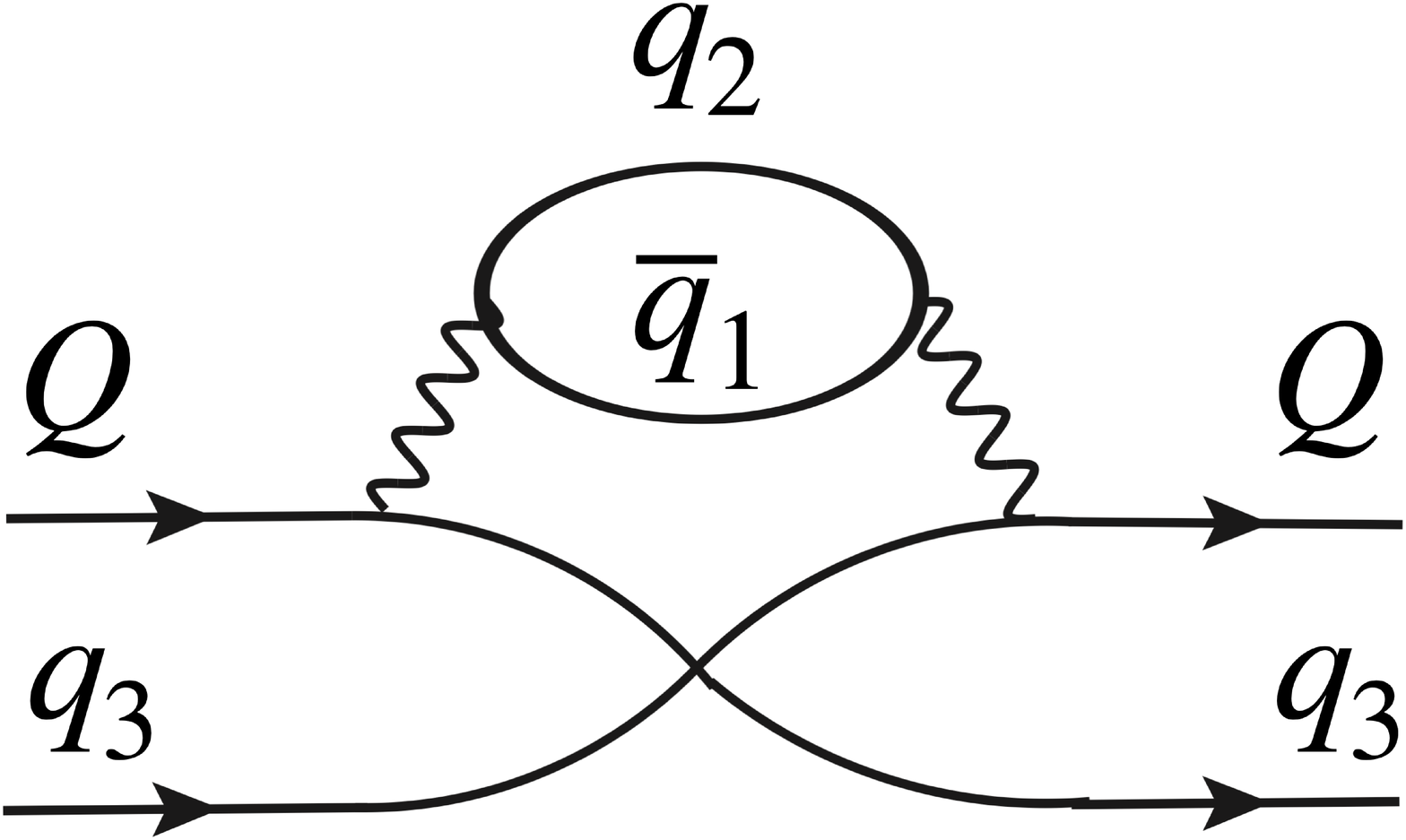}
    \caption{} \label{fig:2c}
  \end{subfigure}
\vspace{-0.5cm}
\caption{
The topological diagrams for the spectator effects: (a) $W$-exchange, (b) destructive (constructive) Pauli interference and (c) constructive (destructive) Pauli interference at the dimension-6 (dimension-7) level. There is {\it no} 4-point vertices between the $W$ bosons and quarks in the diagrams (b) and (c)} 
\label{fig:spectator}
\end{center}
\end{figure}

The dimension-six operator of $\tilde {\cal O}_6$
consists of four types of topologies
\begin{equation}
\tilde {\cal O}_6 = 
\tilde {\cal O}_{we}+\tilde {\cal O}_{int-} + \tilde {\cal O}_{int+}+ \tilde {\cal O}_{\text{SL}},
\end{equation}
where  the subscripts 
$we$, $int+$, $int-$ and SL
stand for $W$-exchange, Pauli constructive interference, Pauli destructive interference and semileptonic, respectively. 
Their topological diagrams  without QCD corrections  are depicted in Fig.~\ref{fig:spectator}, where $\tilde {\cal O}_{int+}$ and $\tilde {\cal O}_{\text{SL}}$ share the same topology, i.e. Fig.~\ref{fig:spectator}(c).
 Four-quark operators for $W$-exchange are decomposed according to their spinor structure
\begin{eqnarray}
\tilde {\cal O}_{we} &=& {\cal C}_{we}^{V-A} 
(Q_\alpha^\dagger L^\mu q_{1\alpha} ) (q_{1\beta}^\dagger L_\mu Q_\beta) 
+\tilde {\cal C}_{we}^{V-A} 
(Q_\beta^\dagger L^\mu q_{1\alpha} ) (q_{1\alpha}^\dagger L_\mu Q_\beta)    \\
&& +{\cal C}_{we}^{S-P} 
(\overline{Q}_\alpha(1+\gamma_5) q_{1\alpha} ) (\overline{q}_{1\beta} (1-\gamma_5) Q_\beta) 
+\tilde {\cal C}_{we}^{S-P} 
(\overline{Q}_\beta(1+\gamma_5) q_{1\alpha} ) (\overline{q}_{1\alpha} (1-\gamma_5) Q_\beta).  \nonumber
\end{eqnarray}
Further, the coefficients are disintegrated as 
\begin{equation}
{\cal C}_{we}^{V-A} = \sum_{i,j}^{1\text{-}6} c_ic_j {\cal K}_{we,ij}^{V-A}\,.
\end{equation}
Thus far we have taken $\tilde{\cal O}_{we}$ and ${\cal C}_{we}^{V-A}$ as an example and the rest of them can be defined in the same manner. 
The explicit expressions of ${\cal K}_{we}$ and ${\cal K}_{int-}$   to the NLO are given in Refs.~\cite{T6_NL} and \cite{Lenz:2013aua} for the nonleptonic and semileptonic decays, respectively, while ${\cal C}_{int+}$ is obtained by the Fierz transformation with massless $q_{1,2,3}$. 

It should be noticed that the full QCD theory and HQET have different coefficients of ${\cal K}$'s  which are related by Eq.~(16) in Ref~\cite{T6_NL}. 
In HQET, one takes the limit of $m_Q\to \infty$ and the series of the HQE is 
truncated to a specific order of $1/m_Q$. In the BM, we do not take the heavy quark limit to  evaluate $I_{{\cal B}_Q}^q$. 
Therefore,  the central values of the computed lifetimes are evaluated with the Wilson coefficients given in the full QCD theory,  whereas  the deviations from the HQET ones are 
treated as the uncertainties.

In the numerical evaluation we have taken the NLO and the Cabibbo suppressed contributions into account. 
As an illustration, we write down the  expressions of the Cabibbo-favored decays to LO. 
We decompose $\langle \tilde {\cal O}_6\rangle $ into several parts
\begin{equation}\label{48}
\frac{G_F^2m_Q^2}{192\pi ^3 } \langle \tilde {\cal O}_6\rangle_{{\cal B}_Q}  = \sum_q \left(
\Gamma_{6,we}^{{\cal B}_Q, q} + \tilde\Gamma_{6,int-}^{{\cal B}_Q, q}  + \Gamma_{6,int-}^{{\cal B}_Q, q} + \Gamma_{6,int+}^{{\cal B}_Q, q}  
+\Gamma_{6,\text{SL}}^{ {\cal B}_Q, q}
\right)\,,
\end{equation}
where $\tilde \Gamma$ contains two massive quarks in the loop integral, while the others without tilde have only one massive quark.
In the cases of $(Q,q_1,q_2,q_3)\in  \{ (c,d,u,s), (b,u,d,c)\}$, $q_1$ and $q_2$ can be taken as massless, and  their expressions are unified as~\cite{Charm:2018,T61,T62}
\bqa \label{eq:T6baryon}
\Gamma_{6,we}^{\B_Q,q_1} &=& {G^2_Fm_Q^2\over 2\pi}\,\xi_Q \,(1-x_Q)^2\Big\{
(c_1^2+c_2^2)\tilde{L}_{{\cal B}_Q} ^{q_1}+2c_1c_2L_{{\cal B}_Q} ^{q_1}\Big\},
 \non \\
\Gamma_{6,int-}^{\B_Q,q_2} &=& -{G_F^2m_Q^2\over 6\pi}\,\xi_Q (1-x_Q)^2\Bigg\{ c_1^2\left[ \left(1+
{x_Q\over 2} \right)\tilde L_{{\cal B}_Q} ^{q_2}-(1+2x_Q) (\tilde S_{{\cal B}_Q} ^{q_2} - \tilde P_{{\cal B}_Q} ^{q_2}) \right]   \non \\
&+& (2c_1c_2+N_cc_2^2)\left[ \left(1+
{x_Q\over 2} \right)L_{{\cal B}_Q} ^{q_2}-
(1+2x_Q )(S_{{\cal B}_Q} ^{q_2} - P_{{\cal B}_Q} ^{q_2})\right]\Bigg\},   \\
\Gamma_{6,int+}^{\B_Q,q_3} &=& -{G_F^2m_Q^2\over 6\pi}\,\xi_Q \Bigg[ 
c_2^2\left( \tilde L_{{\cal B}_Q}^{q_3}-\tilde S_{{\cal B}_Q}^{q_3} + \tilde P_{{\cal B}_Q}^{q_3}\right) 
+ (2c_1c_2+N_cc_1^2)\Big(  L_{{\cal B}_Q}^{q_3}- S_{{\cal B}_Q}^{q_3} +  P_{{\cal B}_Q}^{q_3} \Big) \Bigg] \,, \non 
\eqa	
with $\xi_b = |V_{cb}V^*_{ud}|^2$ and  $\xi_c = |V_{cs}^*V_{ud}|^2$.
We note that $I_{{\cal B}_Q}^q =0 $ if ${\cal B}_Q$ does not contain the quark flavor $q$ as we do not consider the  eye contractions~\cite{Lenz:2013aua}, resulting in  $\Gamma_{6,int-}^{{\cal B}_b,c} =0 $. For singly heavy baryon systems, constructive Pauli interference exists only in charmed baryon decays with $c\to u\bar ds$. 
Since the strange quark does not participate in the quark loop, $x_c$ is absent in $\Gamma_{6,int+}^{\B_c,s}$.

The relevant Wilson coefficient of $\Gamma_{6,int-}^{\B_Q,q_2}$ has the expression $(2c_1c_2+N_cc_2^2-c_1^2)$ which is negative in both charm and bottom sectors. Since $(1+
{x_Q\over 2})L_{{\cal B}_Q} ^{q_2}-
(1+2x_Q )(S_{{\cal B}_Q} ^{q_2} - P_{{\cal B}_Q} ^{q_2})$ is negative, it is evident that the Pauli interference described by $\Gamma_{6,int-}^{\B_Q,q_2}$ is destructive. By the same token,  the Pauli interference described by $\Gamma_{6,int+}^{\B_Q,q_3}$ is constructive since the Wilson coefficient $(2c_1c_2+N_cc_1^2-c_2^2)$ is positive.  

For the process of $(Q,q_1,q_2,q_3) = (b,c,s,c)$, 
the $W$-exchange and  constructive Pauli interference contributions vanish due to the absence of a charm quark in ${\cal B}_b$. It is necessary to consider
the quark masses of both $q_1$ and $q_3$ in the loop integrals, 
and the remaining term comes from  Fig.~\ref{fig:spectator}(b), given as~\cite{Charm:2018}
\begin{eqnarray}
\tilde\Gamma_{6,int-}^{ \B_b,s}&=& - {G_F^2m_b^2\over 6\pi}\,|V_{cb}V_{cs}^*|^2 \sqrt{1-4x_b }\,\Bigg\{
c_1^2\,\Big[ (1-x_b ) \tilde L_{{\cal B}_b}^{s}-(1+2x_b)(\tilde S_{{\cal B}_b} ^{s} - \tilde P_{{\cal B}_b} ^{s} ) \Big]   \non \\
&+& (2c_1c_2+N_cc_2^2)\,\Big[ (1-x_b)L_{{\cal B}_b} ^{s}-
(1+2x_b)(S_{{\cal B}_b} ^{s} - P_{{\cal B}_b} ^{s})\Big]\Bigg\}\,.
\end{eqnarray}
Notice that if we take $x_b=0$, $\tilde \Gamma_{6,int-}^{\B_b,s}$ will be reduced to $\Gamma_{6,int-}^{\B_b,d}$ as it should be since they share the same topological diagram.
A unified expression of $\Gamma_{6,int\pm}^{{\cal B}_Q,q}$  and 
$\tilde \Gamma_{6,int\pm}^{{\cal B}_Q,q}$
can be found in Appendix B of Ref.~\cite{Charm:2022}.

The dimension-7 operators share the same topological diagrams with the dimension-6 ones,
but the NLO correstions are currently still absent. 
In analogy to Eq.~\eqref{48}, we decompose their contributions as~\cite{T71} 
\begin{equation}
\frac{G_F^2m_Q }{192\pi ^3 } \langle \tilde O_7 \rangle _{{\cal B}_Q}  = \sum_q \left(
\Gamma_{7,we}^{{\cal B}_Q, q} + \Gamma_{7,int+}^{{\cal B}_Q, q} + \Gamma_{7,int-}^{{\cal B}_Q, q} + \tilde\Gamma_{7,int-}^{{\cal B}_Q, q} 
\right) \,.
\end{equation}
 Contrary to the previous dimension-6 case, 
the subscripts $int+$ and $int-$ here are referred to Figs. 2(b) and 2(c), respectively. 
We find~\cite{Lenz:2013aua,Charm:2018}
\bqa \label{eq:T7baryon}
\Gamma_{7,we}^{\B_Q,q_1} &=& {G^2_Fm_Q \over 2\pi}\,\xi_Q\,(1-x^2_Q)\Bigg[ 4 c_1c_2 P_3^{q_1} + 2 (c_1^2+c_2^2)\tilde P_3^{q_1}\Bigg],
\non\\
\Gamma_{7,int+}^{\B_Q,q_2} &=& {G_F^2m_Q \over 6\pi}\,\xi_Q \Bigg\{ \Big(2c_1c_2+N_cc_2^2\Big)\Big[-(1-x_Q)^2(1+2x_Q)(P_1^{q_2}+P_2^{q_2}) \non \\
&+& 2(1-x_Q^3)P_3^{q_2}
-12x_Q^2(1-x_Q)P_4^{q_2} \Big]+ c_1^2\Big[ -(1-x_Q)^2(1+2x_Q)(\tilde P_1^{q_2}+\tilde P_2^{q_2})  \non \\
 &+&
 2(1-x_Q^3)\tilde P_3^{q_2}-12(1-x_Q) x_Q^2\tilde P_4^{q_2} \Big]
\Bigg\},   \\
\Gamma_{7,int- }^{\B_Q,q_3} &=& {G_F^2m_Q \over 6\pi}\,\xi_Q\Bigg[ \left(2c_1c_2+N_cc_1^2\right)\Big(-P_1^{q_3}-P_2^{q_3}
+ 2P_3^{q_3} \Big) + c_2^2\Big( -\tilde P_1^{q_3}-\tilde P_2^{q_3}
+ 2\tilde P_3^{q_3} \Big) 
\Bigg]\,. \non
\eqa
for $(Q,q_1,q_2,q_3)\in  \{ (c,d,u,s), (b,u,d,c)\}$ and 
\bqa \label{72}
 \tilde \Gamma_{7,int+}^{\B_b,s} &=& {G_F^2m_b\over 6\pi}\,|V_{cb}V_{cs}^*|^2 \sqrt{1-4x_b}\Bigg\{  \left(2c_1c_2+N_cc_2^2\right)\Big[ -(1+2x_b)(P_1^{s}+P_2^{s}) \non \\
&+& {2\over 1-4x_b}(1-2x_b-2x_b^2)P_3^{s} - {24x_b^2\over 1-4x_b}P_4^{s}  \Big] + c_1^2 \Big[ -(1+2x_b)(\tilde P_1^{s}+\tilde P_2^{s})
\non \\
&+& {2\over 1-4x_b}(1-2x_b-2x_b^2)\tilde P_3^{s}
-{24x_b^2\over 1-4x_b }\tilde P_4^{s}
\Big] \Bigg\}\,,
\eqa
for $(Q,q_1,q_2,q_3) = (b,c,s,c)$. 
In Eqs.~\eqref{eq:T7baryon} and \eqref{72}, $P_{1,2,3,4}^{q}$ are understood as $\langle P_{1,2,3,4}^{q}\rangle_{{\cal B}_Q}$ instead of dimension-7 operators for simplicity. 
The discussions are parallel to the dimension-6 ones. By putting $x_b=0$, we find that $\tilde\Gamma_{7,int-}^{\B_b,s} = \Gamma_{7,int-}^{ \B_b,s}$ and that $x_Q$ is absent in $\Gamma_{7,int-}^{ \B_Q,q_3}$ as $q_3$ does not get involved in the loop integral.

Eqs.~\eqref{eq:T7baryon} and \eqref{72} are derived from the full QCD theory. If one alternatively adopts  HQET instead, there will be several extra terms coming from the $1/m_Q$ corrections to the dimension-6 operators~\cite{T6_NL,Charm:2018}. After including them,   the deviations between two theories start at the dimension-8 level.

We next turn to the semileponic inclusive decay widths. Notice that it is possible to have $(q_1,q_2) = (\ell, \nu_e)$ in Fig.~\ref{fig:spectator}(c). Therefore, the  inclusive semileptonic decay widths also receive corrections from  dimension-6 and -7 operators~\cite{Blok:1994cd}. 
Ignoring the small ratio $z = (m_\mu/m_c)^2 \approx 1\%$, 
the matrix elements are obtained by setting $(c_1,c_2,N_c) = (1,0,2)$ in $\Gamma_{6,int+}^{{\cal B}_c,s}$ and $\Gamma_{7,int-}^{{\cal B}_c,s}$\,, given by
\begin{eqnarray}
		&&\Gamma_ {6, \text{SL}}^{{\cal B}_c,s}=-\frac{G_F^2 m_c^2}{3 \pi}\left|V_{c s}\right|^2 \left(
		L_{{\cal B}_c}^s  -S_{{\cal B}_c}^s  + P_{{\cal B}_c}^s 
		\right)\,,\\
		&&\Gamma_ {7, \text{SL}}^{{\cal B}_c,s}= -\frac{G_F^2 m_c}{3 \pi}\left|V_{c s}\right|^2  \left[
		m_s \left(
		S_{{\cal B}_Q} ^q + P_{{\cal B}_Q} ^q 
		\right) -2E_s 
  L _{{\cal B}_Q} ^q 
		\right]\,.\nonumber
	\end{eqnarray}
This completes the investigation in the contributions of  four-quark operators.
We stress again that the NLO contributions of the dimension-6 operators and Cabibbo-suppressed decays have  been taken into account in the numerical evaluation.

\section{Numerical results and discussions}

Since the total inclusive decay widths are proportional to $m_Q^5$ to the leading-order of the $1/m_Q$ expansion, the numerical results are very sensitive to how well the  heavy quark mass is under control. 
Unfortunately, the choice of the heavy quark mass definition is quite arbitrary to the first-order in QCD, and the pole mass definition, with which the formalism is derived, suffers from the divergences due to the infrared renormalon, imposing a minimal ambiguity proportional to $\Lambda_{\text{QCD}}$~\cite{Beneke:2021lkq}.

For the heavy quark mass $m_Q$  in Eq.~\eqref{optical},
we shall take the pole mass\footnote{To the leading-order of $\alpha_s$, $m_b^{\text{pole}}
= 4.70$~GeV corresponds to $m_b^{\text{kin}}
= 4.57$ GeV with $\mu^{\text{cut}}=1$ GeV, which proves to be suitable for the $B$-meson inclusive decays~\cite{Lenz:2022rbq}, see Eq.~(3.1) and the subsequent discussion in Ref.~\cite{Lenz:2022rbq}. }
\begin{equation}\label{kimas}
(m_b ,m_c )_{\text{pole}} = ( 4.70 \pm 0.10  ,~ 1.59 \pm 0.09 )~\text{GeV}\,,
\end{equation}
where the upper and lower bounds correspond to the two-loop and one-loop results. 
Here we do not consider the other mass schemes such as $\overline{\text{MS}}$, kinetic, MSR and $1S$ schemes as they have to be matched with the pole mass one eventually. The deviations among them are mainly generated by truncating the series of the $\alpha_s$ expansion (see Eq.~(2.54) in Ref.~\cite{Charm:2022}, for instance). As we have allowed a wide range for the pole mass, it will cover the corrections to the order of $O(\alpha_s^2)$.\,\footnote{
For example, $(m_c^{\overline{\text{MS}}},m_c^{{\text{kin}}},m_c^{{\text{MSR}}})
=(1.28, 1.40, 1.36)~$GeV correspond to $m_c^{\text{pole}} = (1.52, 1.67, 1.49)$~GeV to the 
first-order correction in $\alpha_s$, where we have used $\alpha_s(m_c)  = 0.38$ and $\mu^{\text{cut}} = 1$~GeV for the kinetic and MSR mass schemes. For detailed discussions, see Sec. 2.4 of Ref.~\cite{Charm:2022}.
} 
On the other hand, for the quark masses appearing in the loop integral, we fix them with the $\overline{\text{MS}}$ scheme~\cite{FermilabLattice:2018est}
\begin{equation}
\left(\overline{m}_c(\overline{m}_c) ,\overline{m}_s(2~\text{GeV})  \right)_{\overline{\text{MS}}} = (1.3 ,~0.09 )~\text{GeV}\,.
\end{equation}
For the ratio $m_\ell / m_Q$ in the loop integral we take $m_Q$ as the ones in Eq.~\eqref{kimas}.

To be consistent with the NLO corrections in $\Gamma_6$, we use the LO values of the Wilson coefficients from Tables VII and XIII of Ref.~\cite{Buchalla:1995vs}
\begin{eqnarray}
(c_1, c_2)  = ( 1.298, -0.565)\,,
\end{eqnarray}
at $\mu_c= 1.5~\text{GeV}$
for charm quark decays, and 
\begin{eqnarray}
(c_1, c_2,c_3,c_4,c_5,c_6)  = ( 1.139, -0.301,0.013,-0.030,0.009,-0.038) \,,
\end{eqnarray}
at $\mu_b= 4.4~\text{GeV}$
for  bottom quark decays. 
The calculated results of heavy baryon decay widths are summarized in Table~\ref{scen1},
where $\Gamma_3$ stands for the two-quark operator contributions with $\rho_D^3=0$ and $\Gamma_\rho$ is the decay width contributed from the Darwin operator. The semileptonic inclusive branching fractions are defined by 
\begin{equation}\label{SL}
{\cal BF} ^{\text{SL}}_\ell \equiv \tau({\cal B}_Q) 
\Gamma( {\cal B}_Q \to X \ell^\pm\nu_\ell ) 
\,,
\end{equation}
with $\ell = e, \mu , \tau $. 
Since the NLO corrections to $\Gamma_7$ are still absent currently, we shall use the LO value of $\Gamma_7$ for  $\tau$
at the NLO. In the table, the uncertainties arising from $m_Q$, $\mu_H$ and $I_{{\cal B}_Q}^q$ are denoted by the subscripts $m$, $\mu$ and $4$, respectively. The central values are evaluated with full QCD operators, and the deviations from HQET ones are treated as uncertainties denoted by  the subscript $s$.  
Error analyses are not provided for $\Gamma_{6,7}$ in the table for simplicity as their dependence on $m_Q$ is rather weak . The errors from $\mu_H$ and $I_{{\cal B}_Q}^q$ will cause around 11\% and 10\% uncertainties to $\Gamma_{6,7}$ and $\rho_D^3$, respectively. As $\rho_D^3$ is proportional to $\alpha_s$, the Darwin operator does not contribute at the LO and its effect is negligible compared to other hadronic uncertainties at the NLO.

\begin{sidewaystable}
	\caption{ Results for the lifetimes of heavy baryons, where the decay widths and lifetimes are in units of $10^{-12}$ ($10^{-13}$) GeV and $10^{-13}s$ ($10^{-12}s$ ), respectively, for ${\cal B}_c$ (${\cal B}_b$). Uncertainties arising from $m_Q$, $\mu_H$, $I_{{\cal B}_Q}^q$ and the deviation of full QCD from HQET are denoted by the subscripts $m$, $\mu$, $4$ and $s$, respectively. 
In the table, NLO in the first column stands for the numerical results to the NLO precision.
 The NLO values of $\Gamma_7$ are taken to be the same as the ones at the LO}\label{scen1}
		\vspace{6pt}
		\begin{tabular}{lc| ccccccccc|c}\hline 
\multicolumn{2}{c|}{
${\cal B}_Q$ } & $\Gamma_3^{\rm NL} $& $\Gamma_3^{\rm SL} $ & $\Gamma_\rho$ & $\Gamma_{6}^{\text{NL}}$ 
& $\Gamma_6^{\text{SL}}$& $\Gamma_7^{\text{NL} } $
&$\Gamma_7^{\text{SL}}$
&${\cal BF}^{\text{SL}}_e(\%) $&$\tau$ &$\tau_{\text{exp}}$ \\
\hline
\multirow{2}{*}{
$ \Lambda_c^+$ } &LO&  $ 0.85(29)_m $&$ 0.40(13)_m $&$ 0 $&$ 0.75 $&$ 0.01 $&$ 0.49 $&$ 0 $&$ 8.25(78)_m (44)_\mu(37)_4(37)_s$&$ 2.63(46)_m(15)_\mu (12)_4(11)_s$&
\multirow{2}{*}{
$2.029(11)$
}
\\
&NLO&   $ 1.27(42)_m $&$ 0.35(11)_m $&$ 0.07 $&$ 1.26 $&$ 0.01 $&$ 0.49 $&$ 0 $&$ 4.57(42)_m(24)_\mu(21)_4 (13)_s$&$ 1.92(34)_m(11)_\mu (10)_4(5)_s$\\
\hline
\multirow{2}{*}{$ \Xi_c^0 $ }&LO &$ 0.86(28)_m $&$ 0.40(14)_m $&$ 0 $&$ 1.74 $&$ 0.36 $&$ 0.22 $&$ -0.15 $&$ 8.99(58)_m(29)_\mu(25)_4(43)_s $&$ 1.92(31)_m(14)_\mu (12)_4(7)_s$&
\multirow{2}{*}{
$1.505(19)$
}
\\&NLO &
 $ 1.27 (42)_m$&$ 0.35(12)_m $&$ 0.07 $&$ 2.01 $&$ 0.18 $&$ 0.22 $&$ -0.15 $&$ 4.40(45)_m(22)_\mu(19)_4(30)_s $&$ 1.66 (28)_m(11)_\mu (9)_4(6)_s$\\
\hline
\multirow{2}{*}{ $ \Xi_c^+ $ }&LO&
$ 0.86(28)_m $&$ 0.40(14)_m $&$ 0 $&$ 0.26 $&$ 0.35 $&$ -0.09 $&$ -0.15 $&$ 18.59 (26)_m(22)_\mu(19)_4(39)_s$&$ 4.04 (92)_m(10)_\mu (9)_4 (12)_s$&
\multirow{2}{*}{
$4.53(5)$
}
\\&NLO
 & $ 1.27(42)_m $&$ 0.35(12)_m $&$ 0.07 $&$ 0.38 $&$ 0.18 $&$ -0.09 $&$ -0.15 $&$ 8.57 (20)_m(5)_\mu(5)_4(44)_s$&$ 3.27(75)_m(7)_\mu (6)_4(6)_s$\\
\hline
\multirow{2}{*}{$ \Omega_c^0 $ }&LO  & $ 0.91(30)_m $&$ 0.42(14)_m $&$ 0 $&$ 2.34 $&$ 1.22 $&$ -1.09 $&$ -0.83 $&$ 13.51(42)_m (10)_\mu(8)_4(23)_s$&$ 2.22(44)_m(14)_\mu (12)_4(1)_s$& \multirow{2}{*}{
$2.73(12)$
}\\&  NLO
&$ 1.34 (44)_m$&$ 0.37 (12)_m$&$ 0.11 $&$ 2.37 $&$ 0.61 $&$ -1.09 $&$ -0.83 $&$ 1.88(1.33)_m (47)_\mu(40)_4(85)_s$&$ 2.30(51)_m (10)_\mu (9)_4(24)_s$\\
\hline
\hline
\multirow{2}{*}{
$ \Lambda_b$ } &LO& $ 2.28(33)_m $&$ 1.67 (18)_m $&$ 0 $&$ 0.07 $&$ 0 $&$ 0.02 $&$ 0 $&$ 12.34(3)_m (1)_\mu (1)_4(6)_s$&$ 1.63(15)_m (1)_\mu(0)_4(1)_s$&
\multirow{2}{*}{
$1.471(9)$
}
\\
&NLO &  $ 2.78(42)_m $&$ 1.56(17)_m $&$ -0.02 $&$ 0.11 $&$ 0 $&$ 0.02 $&$ 0 $&$ 9.90(3)_m(3)_\mu (3)_4 (3)_s$&$ 1.48(22)_m (1)_\mu(0)_4(1)_s$\\
\hline
\multirow{2}{*}{$ \Xi_b^0 $ }&LO & $ 2.28(33)_m $&$ 1.67(18)_m $&$ 0 $&$ 0.07 $&$ 0 $&$ 0.01 $&$ 0 $&$ 12.37(3)_m(0)_\mu (0)_4(6)_s$&$ 1.63(15)_m (1)_\mu(1)_4(1)_s$&
\multirow{2}{*}{
$1.480(30)$
}
\\&NLO
& $ 2.78(41)_m $&$ 1.56(17)_m $&$ -0.02 $&$ 0.11 $&$ 0 $&$ 0.01 $&$ 0 $&$ 9.94(3)_m(2)_\mu (2)_4(4)_s$&$ 1.49(22)_m(0)_\mu(0)_4(0)_s $\\
\hline
\multirow{2}{*}{ $ \Xi_b^- $ }&LO&
$ 2.28 (33)_m $&$ 1.67(18)_m $&$ 0 $&$ -0.09 $&$ 0 $&$ 0 $&$ 0 $&$ 12.91 (9)_m(6)_\mu(6)_4(6)_s$&$ 1.70 (27)_m(1)_\mu(1)_4(1)_s$&
\multirow{2}{*}{
$1.572(40)$
}
\\&NLO
&   $ 2.78(41)_m $&$ 1.56(17)_m $&$ -0.02 $&$ -0.07 $&$ 0 $&$ 0 $&$ 0 $&$ 10.38(8)_m(2)_\mu(2)_4 (3)_s$&$ 1.55 (23)_m(1)_\mu(0)_4(1)_s $\\
\hline
\multirow{2}{*}{$ \Omega_b^- $ }&LO & $ 2.28(33)_m $&$ 1.67(18)_m $&$ 0 $&$ -0.17 $&$ 0 $&$ -0.04 $&$ 0 $&$ 13.38(15)_m(11)_\mu(10)_4(9)_s $&$ 1.76(28)_m(2)_\mu(2)_4 (1)_s$
& \multirow{2}{*}{
$1.64^{+0.18}_{-0.17}$
}\\&  NLO
&$ 2.77 (41)_m $&$ 1.55(16)_m $&$ -0.03 $&$ -0.15 $&$ 0 $&$ -0.04 $&$ 0 $&$ 10.76 (11)_m(6)_\mu(5)_4(4)_s$&$ 1.60 (25)_m (1)_\mu (0)_4(1)_s$\\
	\hline
		\hline		
	\end{tabular}
\end{sidewaystable}

We see that the differences among the ${\cal B}_Q$'s are mostly ascribed to the four-quark  operators. In particular, $\Gamma_3$ is the same for all antitriplet baryons $T_Q$. 
For ${\cal B}_c$,  the NLO corrections are roughly  50$\%$ in both $\Gamma_3^{\text{NL}}$ and $\Gamma_6^{\text{NL}}$. Assuming the $\alpha_s$ expansion behaves like a geometric series, there will be  roughly  33$\%$ deviations in the lifetimes once the $\alpha_s$ corrections are included to all orders. 
On the other hand, the NLO corrections are found to be around 20$\%$ in ${\cal B}_b$.   From Table \ref{scen1} we see that all the predicted lifetimes are improved to the NLO except for $\tau(\Xi_c^+)$. 
A good sign of the $\alpha_s$ expansion is that
the uncertainties caused by
$\mu_H$ are systematically reduced once the NLO corrections are included. 

It is evident from Table \ref{scen1} that $\Gamma_7^{\rm SL}$ contributes destructively to the total semileptonic rate $\Gamma^{\rm SL}$, while $\Gamma_7^{\rm NL}$ contributes constructively to $\Gamma^{\rm NL}$ for $\Lambda_c^+$ and $\Xi_c^0$ and destructively for $\Omega_c^0$ and $\Xi_c^+$. Owing to the presence of two strange quarks in the $\Omega_c^0$, $\Gamma_7^{\rm SL}(\Omega_c^0)$ and
$\Gamma_7^{\rm NL}(\Omega_c^0)$ are the largest in magnitude among the charmed baryons. Consequently, the lifetime hierarchy pattern $\tau(\Xi_c^+)>\tau(\Lambda_c^+)>\tau(\Xi_c^0)>\tau(\Omega_c^0)$ expected
in HQE to order $1/m_c^3$ is modified to the new one $\tau(\Xi_c^+)>\tau(\Omega_c^0)>\tau(\Lambda_c^+)>\tau(\Xi_c^0)$ in HQE to order $1/m_c^4$.
The $\Omega_c^0$ could live longer than the $\Lambda_c^+$ due to the suppression from $1/m_c$ corrections arising from dimension-7 four-quark operators.

The $1/m_c$ expansion in the matrix elements of four-quark operators seems to work reasonably well for the antitriplet charmed baryon $T_c$ where $\Gamma_7$ is around $40\%$ compared to $\Gamma_6$. 
However, the $1/m_c$ expansion could be problematic for the $\Omega_c$ if $\Gamma_7^{\text{SL}}({\rm NLO})$  
is larger than $\Gamma_6^{\text{SL}}({\rm NLO})$ in magnitude, leading to a smaller or even negative semileptonic rate $\Gamma^{\text{SL}}(\Omega_c^0)$. Also recall that the deviation of the full QCD from HQET  is huge for the $\Omega_c$.
A future study of NLO corrections to $\Gamma_7$ and dimension-8 operators  will be useful to clarify the issue. 
Although the computed results for $\tau({\cal B}_c)$ are compatible with the current data, we need to keep in mind the possible shortcomings in the higher-order  $1/m_c$ expansion.

\begin{table}[t!]
\caption{
Comparison of our results with Ref. \cite{Charm:2022} obtained in the pole mass scheme for charmed baryons and Ref. \cite{bottom:2023} in the kinetic mass scheme for bottom baryons, of which the uncertainties are added quadratically. The baryon matrix elements in Refs. \cite{Charm:2022,bottom:2023} are evaluated using the NRQM. Experimental results are quoted from Ref. \cite{PDG2022} and Table \ref{tab:expt_lifetimes}. The lifetimes are in units of $10^{-13}s$ for ${\cal B}_c$ and $10^{-12}s$ for ${\cal B}_b$}\label{scen2}
		\begin{tabular}{c| cc|cc|cc}\hline 
\hline
& \multicolumn{2}{|c|}{BM\,\footnote{
~Corresponding to the results in which the CMM is removed from the bag model. 
   }}
   & \multicolumn{2}{|c|}{NRQM}
   & \multicolumn{2}{|c}{Experiment} \\
  \hline
  \hline
~~${\cal B}_Q$~~
&${\cal BF}^{\text{SL}}_e(\%) $&$\tau$ &${\cal BF}^{\text{SL}}_e(\%) $&$\tau$
& ${\cal BF}^{\text{SL}}_e(\%) $&
$\tau$ \\
\hline
$ \Lambda_c^+$   &$4.57(54)$&$ 1.92(37) $ & $3.80^{+0.58}_{-0.57}$ & $ 3.04^{+1.06}_{-0.80} $ & 
$3.95\pm0.35$ & $2.029(11)$
\\
\hline
$ \Xi_c^0 $ &$ 4.40(61) $&$ 1.66(32)$ 
&$4.31^{+0.87}_{-0.84}$
&
$2.31^{+0.84}_{-0.59}$
& 
-
&
$1.505(19)$\\
\hline
 $ \Xi_c^+ $ &$ 8.57(49) $&$ 3.27 (76)$ & 
$12.74^{+2.54}_{-2.45}$&$4.25^{+1.22}_{-1.00}$ &
- &
$4.53(5)$\\
\hline
$ \Omega_c^0 $ 
&$1.88(1.69)  $&$ 2.30(58) $
&$7.59^{+2.49}_{-2.24}$
&$2.59^{+1.03}_{-0.70}$
&-
&$2.73(12)$
\\
\hline
\hline
$ \Lambda_b$ &$ 9.90(3) $&$ 1.48(22)$ &  
$11.0^{+0.6}_{-0.5}$
&$1.490 _{- 0.207 }^{+ 0.176 }$ & 
-&
$1.471(9)$\\
\hline
$ \Xi_b^0 $ &$ 9.94(6) $&$ 1.49(22)$ 
&  
$11.1^{+0.6}_{-0.6}$ 
&$1.493 _{- 0.207 }^{+ 0.177 }$&-
&
$1.480(30)$\\
\hline
$ \Xi_b^- $&$ 10.38(9)$&$ 1.55(23)$ &  
$11.7^{+0.7}_{-0.6}$
&$1.608 _{- 0.230 }^{+ 0.194 }$ &-
& $1.572(40)$
\\
\hline
$ \Omega_b^- $ &$ 10.76(14)$&$ 1.60 (25)$&
$12.0^{+1.4}_{-1.4}$ 
&$1.692 _{- 0.261 }^{+ 0.231 }$&-
& $1.64^{+0.18}_{-0.17}$\\
	\hline
		\hline		
	\end{tabular}
\end{table} 

As for the bottom baryon ${\cal B}_b$, HQE works nicely, obeying the hierarchy\footnote{An exception to the hierarchy is that $\Gamma_\rho < \Gamma_7$. } $\Gamma_3 \gg \Gamma_6 > \Gamma_7$.
Nonetheless,  $m_b\gg \Lambda_{\text{QCD}}$ is a double blade. Although it ensures the validity of HQE, the numerical results are extremely sensitive to $m_Q$. 
If we alternatively take $\tau_{\text{exp}}(B) $ as input to fix $m_b$ (see the footnote of Eq.~\eqref{kimas}), then the uncertainties denoted by the subscript $m$  can be discarded in Table~\ref{scen1} for ${\cal B}_b$.   
To acquire the predictive power, we compute the lifetime ratios where the $m_Q$ dependence is largely canceled. 
We define 
$\tau_\Delta ({\cal B}_b) \equiv 1-  \tau({\cal B}_b)  / \tau( \Lambda_b) $ and find that
\begin{eqnarray}
\left(
\tau_\Delta(\Xi_b^0) ,\tau_\Delta(\Xi_b^-), \tau_\Delta(\Omega_b^-)  
\right) = \begin{cases}
\left(
0.2\pm 2.1, 7.8\pm 2.1  , 13.2\pm 4.7 
\right)\%,   & \quad \mbox{NRQM}   \\
\left(
0.38\pm 0.10, 4.90\pm 1.28  , 8.35 \pm 2.22 
\right)\%,  & \quad \mbox{BM} \\
\left(
0.6\pm 2.1, 6.9\pm 2.8  , 11.5 ^{+ 12.2}_{-11.6} 
\right)\%,  & \quad \mbox{Exp}  \end{cases}
\end{eqnarray}
where the predictions of NRQCD are quoted from Ref.~\cite{bottom:2023}. 
We see that while the lifetime pattern $\tau(\Omega_b^-)> \tau(\Xi_b^-) > \tau(\Xi_b^0) > \tau(\Lambda_b^0)$ is respected by theory and partially by experiment, 
we predict smaller values for 
$\tau_\Delta(\Xi_b^-)$ and $ \tau_\Delta(\Omega_b^-)$ compared to the NRQM due to the difference in the ratio $I_{{\cal B}_b} ^s/ I_{{\cal B}_b} ^{u,d} $ and the treatment of $C_G(m_Q,\mu)$.

To compare the BM and NRQM, we collect the calculated ${\cal BF}_e^{\text{SL}}$ and $\tau$ in Table~\ref{scen2}. 
 The numerical values of $\Lambda_c^+$
in both models
are consistent with experiment. Nevertheless, the results of ${\cal BF}_e^{\text{SL}}(\Xi_c^+)$ and ${\cal BF}_e^{\text{SL}}(\Omega_c^0)$ deviate largely. Notice that if we focus only on the 
Cabibbo-favored semileptonic decays of $c\to s e^+ \nu_e$, then we will have 
\begin{equation}
\frac{ {\cal BF}_e^{\text{SL}}(\Xi_c^+)}{ \tau(\Xi_c^+)  } =
\frac{ {\cal BF}_e^{\text{SL}}(\Xi_c^0)}{ \tau(\Xi_c^0)  }\,,
\end{equation}
under the isospin symmetry, which is well respected by the BM.
As for ${\cal BF} _e ^{\text{SL}}(\Omega_c^0)$, the difference stems from the treatment for dimension-7 operators. If we take $E_s = 0.33$ GeV instead of $0.5$ GeV (see the discussions below Eq.~\eqref{approx}), our value will turn out to be compatible with Ref.~\cite{Charm:2022}. Conversely, if we take $\Lambda_{\text{QCD}} = 0.5$ GeV for the NRQM, the LO value of $ \Gamma_7^{\text{SL}}$ will be larger than $ \Gamma_6^{\text{SL}}$ in magnitude in Table 24 of Ref.~\cite{Charm:2022}, resulting in a much smaller value of ${\cal BF}_e^{\text{SL}}(\Omega_c^0)$.

As a final remark, we point out that $\Xi_Q-\Xi_Q'$ mixing is of the order of $m_s/m_Q$. It will induce a nonzero $\mu_G^2$ in $\Xi_Q$~\cite{Matsui:2020wcc} but its effect is beyond the scope of this work. 

\section{Conclusion}

We have studied the inclusive decay widths of singly heavy baryons with the final results summarized in Table~\ref{scen1}. The nonperturbative baryon matrix elements are estimated using the improved bag model in which the unwanted CMM has been removed. 
 The running of $\mu_G^2$ and $I_{{\cal B}_Q}^q$ was discussed under the full QCD theory and HQET. To the leading-order of $\alpha_s$, we found that $\tilde B=1$ holds irrespective of the energy scale in the $\overline{\text{MS}}$ scheme of the full QCD theory. 
The results of $\mu_\pi^2$, $\mu_G^2$ and $\rho_D^3$ are in good agreement with the literature
but a large discrepancy is found in $I_{{\cal B}_b}^q$. 
The requirement of $I_{{\cal B}_b}^q\approx I_{{\cal B}_c}^q$ from the heavy quark limit
holds nicely in the BM but is badly broken in the NRQM.
In particular, $L_{\Omega_b}^s$ obtained in the latter is nearly four times larger than the one in the former. 
As a result, we have $\tau_\Delta(\Omega_b^-) = (8.34 \pm 2.22)\%$ in contrast to $(13.2 \pm 4.7)\% $ obtained in the NRQM, while the current data lead to $(11.5^{+12.2}_{-11.6})\% $. 

 We found an excellent agreement between theory and experiment for the lifetimes of bottom baryons even at the dimension-6 level. Effects of dimension-7 operators are rather small.
As for charmed baryons, the calculated lifetimes are consistent with the current experiments and we found that the established  new hierarchy $\tau(\Xi_c^+)>{\tau(\Omega_c^0)}>\tau(\Lambda_c^+)>\tau(\Xi_c^0)$ is traced back to the destructive contributions 
from the dimension-7 operators in the $\Omega_c^0$. This confirms the speculation made in Ref.~\cite{Charm:2018}, namely, the $\Omega_c^0$, which is naively expected to be shortest-lived in the charmed baryon system owing to the large constructive Pauli interference, could live longer than the $\Lambda_c^+$ due to the suppression from $1/m_c$ corrections arising from dimension-7 four-quark operators.

To discriminate between the BM and NRQM approaches, 
we recommend to measure ${\cal BF}^{\text{SL}}_e(\Xi_c^+)$
and
${\cal BF}^{\text{SL}}_e(\Omega_c^0)$ in the forthcoming experiments as the BM and NRQM differ significantly mainly due to the treatment of dimension-7 operators. 
A possible sign for the failure of HQE in the $\Omega_c^0$ will occur if $\Gamma_7^{\text{SL}}({\rm NLO})$  is larger than $\Gamma_6^{\text{SL}}({\rm NLO})$ in magnitude but opposite in sign, leading to a smaller or even negative semileptonic rate $\Gamma^{\text{SL}}(\Omega_c^0)$.
 Hence, a study of $\Gamma_7$ at the NLO and  dimension-8 operators is  needed to settle down the issue.

	\begin{acknowledgments}
		This research was supported in part by the Ministry of Science and Technology of R.O.C. under Grant No. MOST-111-2112-M-001-036 and 
		the National Natural Science Foundation of China
		under Grant No. 12205063.
	\end{acknowledgments}

\appendix 
\section{Four-quark operator matrix elements in the bag model}

In this appendix, we sketch the method of evaluating  $I_{{\cal B}_Q}^q$ in the BM with the center-of-mass motion removed.
To be consistent with HQE, the normalization constant in the BM is chosen to satisfy~\cite{Liu:2022pdk}
\begin{equation}\label{normalization}
\langle {\cal B}_Q, p^\mu | {\cal B}_Q, p^{\prime \mu} \rangle = 2 p^0 (2\pi)^3  \delta^3( \vec{p} - \vec{p}')\,,
\end{equation}
where $p^{(\prime)}$ is the 4-momentum of the baryons. To get the correct normalization constant, boosting the states at rest is required. 
The readers interested in the technical details are referred to Ref.~\cite{Liu:2022pdk}. Here, we simply quote the results 
\begin{equation}\label{overlap}
\frac{1}{\mathcal{N} ^2}=\frac{1}{2M_{{\cal B}_Q}} \int d^3 \vec{x}_{\Delta} \prod_{i=1,2,3} D_{q_i}\left(\vec{x}_{\Delta}\right)\,, 
\end{equation}
where $q_i$ stands for the $i$-th quark and 
\begin{equation}
D_q \left(\vec{x}_{\Delta}\right)=  \int d^3 \vec{x} \phi_q^{\dagger}\left(\vec{x}^+\right) \phi_q\left(\vec{x}^-\right)
=\int d^3 \vec{x}\left[u_q^{+} u_q^{-}+v_q^{+} v_q^{-}\left(\hat{x}^{+} \cdot \hat{x}^{-}\right)\right]\,,
\end{equation}
with $\vec{x}^\pm = \vec{x} \pm \vec{x}_\Delta /2 $ and $\hat{x}^\pm$ the norm of $\vec{x}^\pm$. 
Here, $\phi_q(\vec{x}^\pm ) $ are the static bags centering at $\mp \vec{x}_\Delta/2 $, $D_q(\vec{x}_\Delta)$ describes their overlapping, and the use of the abbreviation 
\begin{equation}
\phi_q \left(\vec{x} ^\pm \right) = \left(
\begin{array}{c}
u^\pm _q \chi \\ 
iv^\pm _q \left( \hat{x}^\pm \cdot \vec{\sigma} \right)  \chi 
\end{array}
\right)\,,
\end{equation}
has been adopted. 

As for the matrix elements, let us start with $L^q_{{\cal B}_Q}$ defined in Eq. (\ref{4quark}).
With the help of the anticommutation relation
\begin{equation}
\{ q_{a\alpha}^\dagger (\vec{x}), q _{b \beta}(\vec{x}') \}  = \delta_{ab} \delta_{\alpha\beta} \delta^3 (\vec{x} - \vec{x}')\,,
\end{equation}
we find that 
\begin{eqnarray}\label{22}
L_{{\cal B}_Q}^q 
&=&
\frac{{\cal N}^2}{ 2 M_{{\cal B}_Q }} 
\sum_{[\lambda]} {\cal F}([\lambda], {\cal B}_Q)  
\int d^3\vec{y} d^3 \vec{y}\,' d^3 \vec{x}_3
\phi^\dagger_{q_3}(\vec{x}_3 - \vec{y}) \phi_{q_3} (\vec{x}_3 - \vec{y}\,' )\\
&&\qquad\qquad\qquad\qquad
\phi_{Q\lambda_4}^\dagger(-\vec{y} ) L^\mu \phi_{q\lambda_3 }(-\vec{y} \, ' ) 
\phi_{q\lambda_2}^\dagger(-\vec{y} ) L^\mu \phi_{Q\lambda_1}(-\vec{y} \, ' )\,,  \nonumber\\
&=&
\frac{{\cal N}^2}{ 2 M_{{\cal B}_Q }} 
\int d^3 \vec{x}_\Delta D_{q_3}(\vec{x}_\Delta)  {\cal L} (\vec{x}_\Delta, {\cal B}_Q )\,,\nonumber
\end{eqnarray}
where 
we have changed the integration variables 
$
\left(\vec{y}, \vec{y}\,^{\prime}, \vec{x}_3\right) \rightarrow\left(-\vec{x}^+,-\vec{x}^-, \vec{x}_3-\vec{x}\right)$,
$q_3$ is the spectator quark, $[\lambda] = (\lambda_1,\lambda_2,\lambda_3,\lambda_4)  $, 
and 
\begin{equation}\label{22b}\small
{\cal L}(\vec{x}_\Delta , {\cal B}_Q) = \sum_{[\lambda]} {\cal F}([\lambda], {\cal B}_Q)  
\int d^3 \vec{x} \phi_{Q \lambda_4}^{\dagger}\left(\vec{x}^{+}\right) L_\mu \phi_{q \lambda_2}\left(\vec{x}^{-}\right) \phi_{q \lambda_3}^{\dagger}\left(\vec{x}^{+}\right) L^\mu \phi_{Q \lambda_1}\left(\vec{x}^{-}\right)\,.
\end{equation}
 The spinor-flavor overlapping ${\cal F}$ reads
 \begin{eqnarray}\label{24}
&&\sum_{[\lambda]}  {\cal F}([\lambda], T_Q) (
\lambda_1 \otimes \lambda_2 \otimes \lambda_3 \otimes \lambda_4
) = \frac{1}{4}\left(
\uparrow \uparrow \uparrow \uparrow + 
\uparrow \downarrow \downarrow \uparrow +
\downarrow  \uparrow  \uparrow \downarrow + 
\downarrow \downarrow\downarrow \downarrow 
\right) \,, \\
&&\sum_{[\lambda]}  {\cal F}([\lambda], \Omega_Q) (
\lambda_1 \otimes \lambda_2 \otimes \lambda_3 \otimes \lambda_4
) =  \frac{1}{6} \left( 
5\downarrow \uparrow \uparrow \downarrow + 5 \uparrow \downarrow \downarrow \uparrow + \uparrow\uparrow\uparrow\uparrow + \downarrow \downarrow\downarrow\downarrow -4 \uparrow\uparrow\downarrow \downarrow - 4 \downarrow \downarrow \uparrow \uparrow
\right) \,, \nonumber 
 \end{eqnarray}
where the baryon spins are traced to simplify the formalism. 

As we have traced over the baryon spins, the spinors cannot depend on specific directions, 
{\it i.e.}
\begin{eqnarray}\label{25}
&&\sum_{[\lambda]}  {\cal F}([\lambda], {\cal B}_Q) (\chi_{\lambda_4}^\dagger \chi _{\lambda_3 } )(\chi_{\lambda_2}^\dagger \chi _{\lambda_1 })   = {\cal C}_{\text{unflip}} ^ {{\cal B}_Q}\,,\nonumber\\
&&\sum_{[\lambda]}  {\cal F}([\lambda], {\cal B}_Q) (\chi_{\lambda_4}^\dagger \sigma_i \chi _{\lambda_3 } )(\chi_{\lambda_2}^\dagger \chi _{\lambda_1 }) 
= \sum_{[\lambda]}  {\cal F}([\lambda], {\cal B}_Q) (\chi_{\lambda_4}^\dagger  \chi _{\lambda_3 } )(\chi_{\lambda_2}^\dagger \sigma_i \chi _{\lambda_1 })  = 0 \,, \nonumber\\
&&\sum_{[\lambda]}  {\cal F}([\lambda], {\cal B}_Q) (\chi_{\lambda_4}^\dagger \sigma_i \chi _{\lambda_3 } )(\chi_{\lambda_2}^\dagger \sigma_j \chi _{\lambda_1 })  
={\cal C}_{\text{flip}} ^ {{\cal B}_Q} \delta_{ij} \,.
\end{eqnarray}
From Eq.~\eqref{24}, it is straightforward to show that
\begin{equation}\label{flipunflip}
({\cal C}_{\text{unflip}}^{T_Q}, {\cal C}_{\text{flip}}^{T_Q}) = (1/2, 1/2)\,,~~~ ({\cal C}_{\text{unflip}}^{\Omega_Q}, {\cal C}_{\text{flip}}^{\Omega_Q}) = (-1, 5/3)\,.     
\end{equation}
Plugging Eq.~\eqref{25} into Eq.~\eqref{22}, we arrive at \begin{equation}\label{col1}
{\cal L} (\vec{x}_\Delta, {\cal B}_Q)  = \int d^3\vec{x} \sum_{k=1,2,3,4} \Upsilon^{{\cal B}_Q}_k(\vec{x}_\Delta,\vec{x})\,,
\end{equation}
where 
	\begin{eqnarray}\label{col2}
\Upsilon_1^{{\cal B}_Q}(\vec{x}_\Delta,\vec{x}) &=&
{\cal C}_{\text{unflip}}^{{{\cal B}_Q}} 	\left( 
	u_Q^+ u_q^- +  v_Q^+v_q^-  
	\hat{x}^+ \cdot \hat{x}^-  
	\right)\left( 
	u_q^+ u_Q^- +  v_q^+v_Q^-  
	\hat{x}^+ \cdot \hat{x}^-   
	\right) \nonumber\\
	&&-{\cal C}_{\text{flip}}^{{{\cal B}_Q}} \frac{(\vec{x}_\Delta \times \vec{x}  ) ^2}{(r^+r^-)^2}  v_Q^+v_q^- v_q^+v_Q^- \,, \nonumber\\
\Upsilon_2^{{\cal B}_Q} (\vec{x}_\Delta,\vec{x}) &=&-{\cal C}_{\text{flip}}^{{{\cal B}_Q}}   \left(   u_Q^+v_q^- \hat{x}^- -  v_Q^+  u_q^-  \hat{x}^+ \right) 
	\left( u_q^+  v_Q^-  \hat{x}^- -  v_q^+  u_Q^-  \hat{x}^+ \right) \,,\nonumber\\
\Upsilon_3^{{\cal B}_Q}  (\vec{x}_\Delta,\vec{x})&=&  -\frac{ {\cal C}_{\text{unflip}}^{{\cal B}_Q} } { {\cal C}^{{\cal B}_Q}_{\text{flip}}}\Gamma_2^{{\cal B}_Q} (\vec{x}_\Delta,\vec{x})
	-2 {\cal C}_{\text{flip}}^{{\cal B}_Q}  \left( u_Q^+  v_q^-  \hat{x}^- +  v_Q^+  u_q^-  \hat{x}^+ \right) \cdot
	\left( u_q^+  v_Q^-  \hat{x}^- +  v_q^+  u_Q^-  \hat{x}^+ \right)  \,,\nonumber\\
\Upsilon_4^{{\cal B}_Q}(\vec{x}_\Delta,\vec{x}) &=&  -{\cal C}_{\text{flip}}^{{\cal B}_Q}   \Big[ 3 u_Q^+u_q^- u_q^+ u_Q^- +
	v_Q^+v_q^- v_q^+ v_Q^-  \left( 2 +(\hat{x}^+ \cdot\hat{x}^-)^2 \right)  \\
	&&-(u_Q^+u_q^- v_q^+ v_Q^- + v_Q^+v_q^- u_q^+ u_Q^-  ) \hat{x}^+\cdot \hat{x}^-\Big]+	 {\cal C}_{\text{unflip}}^{{\cal B}_Q}  v_Q^+ v_q^- v_q^+ v_Q^-  \frac{   (\vec{x}_\Delta \times \vec{x}) ^2 }{(r^+r^-)^2} \,,\nonumber
\end{eqnarray}
with  the abbreviations $r^\pm = |\vec{x} \pm \vec{x}_\Delta /2 |$.
Likewise, for the scalar and pseudo-scalar operators given in Eq. (\ref{4quark}) we have 
\begin{eqnarray}\label{SP}
&&S_{{\cal B}_Q}^q = \frac{{\cal N}^2 } { 2 M_{{\cal B}_Q}}  \int  d^3 \vec{x}_\Delta 
D_{q_3} (\vec{x}_\Delta) {\cal S}(\vec{x}_\Delta, {\cal B}_Q) \,, \nonumber\\
&&P_{{\cal B}_Q}^q = \frac{{\cal N}^2 } { 2 M_{{\cal B}_Q}}  \int  d^3 \vec{x}_\Delta 
D_{q_3} (\vec{x}_\Delta) {\cal P}(\vec{x}_\Delta, {\cal B}_Q)\,,
\end{eqnarray}
where 
\begin{eqnarray}\label{in3}
{\cal S} (\vec{x}_\Delta,{{\cal B}_Q} ) &=&
\int d^3 \vec{x} \Big[ 
{\cal C}_{\text{unflip}}^{{{\cal B}_Q}} 	\left( 
	u_Q^+ u_q^- -  v_Q^+v_q^-  
	\hat{x}^+ \cdot \hat{x}^-  
	\right)\left( 
	u_q^+ u_Q^- -  v_q^+v_Q^-  
	\hat{x}^+ \cdot \hat{x}^-   
	\right)  \nonumber\\
 && \qquad
-{\cal C}_{\text{flip}}^{{{\cal B}_Q}} \frac{(\vec{x}_\Delta \times \vec{x}  ) ^2}{(r^+r^-)^2}  v_Q^+v_q^- v_q^+v_Q^- \Big]\,, \nonumber\\
{\cal P} (\vec{x}_\Delta,{{\cal B}_Q} ) &=& - {\cal C}_{\text{flip}}^{{\cal B}_Q} \int d^3 \vec{x}    \left(   u_Q^+v_q^- \hat{x}^- +  v_Q^+  u_q^-  \hat{x}^+ \right) 
	\left( u_q^+  v_Q^-  \hat{x}^- +  v_q^+  u_Q^-  \hat{x}^+ \right)\,. 
\end{eqnarray}
The correctness of the formalism can be checked by taking $\vec{x}_\Delta = 0$, which will bring us to the static bag results. Taking the scalar and pseudo-scalar operators as an example, we obtain 
\begin{eqnarray}
&&\frac{1}{2M_{T_Q}} \langle T_Q | \overline{Q}(1-\gamma_5) q \overline{q}(1+\gamma_5) Q| T_Q \rangle_{SB} = {\cal S}(0,T_Q) - {\cal P}(0,T_Q)\\
&&= \frac{1}{2}\int d^3\vec{x} \left[ (
u_Qu_q - v_Q v_q )^2
+(u_Q v_q + v_Q u_q) ^2 
\right] = \int d^3 \vec{x} \left(
u_Q^2 u_q^2 + v_Q^2 v_q^2 + u_Q^2 v_q^2 + v_Q^2 u_q^2
\right)\,,  \nonumber
\end{eqnarray}
which is consistent with Eq.~(4.4) of Ref.~\cite{Charm:2018}. 
To examine the heavy-quark limit,
we ignore the small component of the $\phi_Q$, {\it i.e.} $v^\pm_Q = 0$, resulting in 
\begin{eqnarray}
 \lim_{v_Q\to 0}
{\cal L}(\vec{x}_\Delta, {\cal B}_Q) &=& \left( {\cal C}_{\text{unflip}}^{{\cal B}_Q} 
-3 {\cal C}_{\text{flip}}^{{\cal B}_Q} 
\right)
\int d^3\vec{x}
\left( u_Q^+ u_Q^- u_q^+u_q^-  \right)\nonumber\\
&&\qquad
- \left( {\cal C}_{\text{unflip}}^{{\cal B}_Q} 
+ {\cal C}_{\text{flip}}^{{\cal B}_Q} 
\right)
\int d^3\vec{x}
\left( u_Q^+ u_Q^- v_q^+v_q^-  \right) \hat{x}^+ \cdot \hat{x}^-
\,,
\nonumber\\
 \lim_{v_Q\to 0} {\cal S}(\vec{x}_\Delta, {\cal B}_Q) &=&  {\cal C}_{\text{unflip}}^{{\cal B}_Q} 
\int d^3\vec{x}
\left( u_Q^+ u_Q^- u_q^+u_q^-  \right) \,,
\nonumber\\
\lim_{v_q\to 0 }{\cal P}(\vec{x}_\Delta, {\cal B}_Q) &=& - {\cal C}_{\text{flip}}^{{\cal B}_Q} 
\int d^3\vec{x}
\left( u_Q^+ u_Q^- v_q^+v_q^-  \right) \hat{x}^+ \cdot \hat{x}^-
\,,
\end{eqnarray}
leading to the relations
\begin{equation}
L_{{\cal B}_Q}^q = \frac{{\cal C}_{\text{unflip}}^{{\cal B}_Q} 
-3 {\cal C}_{\text{flip}}^{{\cal B}_Q} }{
{\cal C}_{\text{unflip}}^{{\cal B}_Q} 
} S_{{\cal B}_Q}^q  
+\frac{{\cal C}_{\text{unflip}}^{{\cal B}_Q} 
+ {\cal C}_{\text{flip}}^{{\cal B}_Q} }{
{\cal C}_{\text{flip}}^{{\cal B}_Q} 
} P_{{\cal B}_Q}^q+{\cal O}\left( \frac{1}{m_Q}
\right) \,.
\end{equation}
Substituting Eq.~\eqref{flipunflip} into the above equation yields 
\begin{equation}
L_{T_Q} ^q = -2 S_{T_Q}^q +2 P_{T_Q}^q\,,~~~L_{\Omega_Q} ^q = 6 S_{\Omega_Q}^q + \frac{2}{5} P_{\Omega_Q}^q\,. 
\end{equation}
We note that the first equation for $T_Q$
also holds to $O(1/m_Q)$ as one can check by keeping the linear terms of $v^\pm_Q$. 
It is not an accident in the BM but a general result closely related to the Fierz identity shown in Ref.~\cite{Neubert:1996we}. 
Furthermore, the relations in the NRQM can be obtained by taking $v_q = 0 $, resulting a vanishing $P_{{\cal B}_Q}^q$  and 
\begin{equation}
\lim_{m_q\to \infty}
\left(  
\frac{
L_{{ T}_Q}^q
}{
S_{{ T}_Q}^q
} 
\right) 
= -2 \,,~~~\lim_{m_q\to \infty}
\left( 
\frac{
L_{\Omega_Q}^s
}{
S_{\Omega_Q}^s
} \right) 
=\lim_{m_q\to \infty}
\left( 
\frac{
L_{\Omega_Q}^s
}{
L_{\Xi _Q}^s
} \right) 
= 6 
\end{equation}
which are consistent with the literature. 

 To evaluate the $d^3\vec{x}$ integrals in Eqs.~\eqref{overlap}, \eqref{col1} and \eqref{in3}, it is convenient to choose the cylindrical coordinate $\vec{x} = (\rho, \Phi, z)$ with $\vec{x}_\Delta \parallel \vec{z}$. In this coordinate, the scalar products of the vectors are given by
 \begin{equation}
\begin{aligned}
&\hat{x}^+\cdot \hat{x}^- = \frac{ 4 \rho^2 + 4z^2 - r_\Delta^2}{4r^+ r^- }\,,\\
&(\vec{x} \times \vec{x}_\Delta) ^2  = \rho^2 r_\Delta^2 \,,
\end{aligned}
 \end{equation}
 where  $r_\Delta = |\vec{x}_\Delta|$. It is then straightforward to see that the integrands are independent of $\Phi$, yielding 
 \begin{equation}
\int d^3\vec{x}\, {\cal I}(\vec{x}_\Delta, \vec{x}) =2 \pi \int_0^{\sqrt{R^2-r_{\Delta}^2 / 4}} \rho d \rho \int_{-\sqrt{R^2-\rho^2}+r_{\Delta} / 2}^{\sqrt{R^2-\rho^2}-r_{\Delta} / 2} d z {\cal I}(r_\Delta , \rho, z)\,,
 \end{equation}
 where ${\cal I}(\vec{x}_\Delta,\vec{x})$ stands for the integrands in Eqs.~\eqref{overlap}, \eqref{col1} and \eqref{in3}.  
The boundaries of the  integrals come from $\phi_q(r^\pm) = 0$ for $r^\pm >R$, and we have integrated out $\Phi$ in the second equation. The dependence on the direction of $\vec{x}_\Delta$ has been dropped as the integrands depend only on $r_\Delta$. 
Finally, since $D_q(\vec{x}_\Delta)$, ${\cal L}(\vec{x}_\Delta, {\cal B}_Q)$, 
${\cal S}(\vec{x}_\Delta, {\cal B}_Q)$ and ${\cal P}(\vec{x}_\Delta, {\cal B}_Q)$ depend only on the magnitude of $\vec{x}_\Delta$, we can replace $\int d^3\vec{x}_\Delta$ by $4 \pi \int d r_\Delta$ in Eqs.~\eqref{overlap}, \eqref{22}, and \eqref{SP}.  We see that the sextuple integrals~$(d^3\vec{x}_\Delta, d^3\vec{x})$ therein are reduced to the triple ones~$(dr_\Delta,d\rho, dz)$, which largely brings down the computing time in the numerical evaluation. A modern computer shall be able to carry out the integrals in no time.

\end{document}